\documentclass[aps,floatfix,twocolumn,showpacs,pra]{revtex4}
\usepackage{graphicx,bm}

\begin{document}
\title{Non-diffusive phase spreading of a Bose-Einstein condensate at finite temperature}
\author{A. Sinatra}
\affiliation{Laboratoire Kastler Brossel,
Ecole Normale Sup\'erieure,
24 Rue Lhomond, 75231 Paris Cedex 05, France}
\author{E. Witkowska}
\affiliation{Institute of Physics, Polish Academy of Sciences,
Aleja Lotnik\'ow 32/46, 02-668 Warszawa, Poland }
\author{Y. Castin}
\affiliation{Laboratoire Kastler Brossel,
Ecole Normale Sup\'erieure,
24 Rue Lhomond, 75231 Paris Cedex 05, France}

\begin{abstract}
We show that the phase of a condensate in a finite temperature gas
spreads linearly in time at long times rather than in a diffusive way.
This result is supported by classical field simulations, and analytical
calculations which are generalized to the quantum case under the
assumption of quantum ergodicity in the system.
This super-diffusive behavior is intimately related to
conservation of energy during the free evolution of the system and to
fluctuations of energy in the prepared initial state.
\end{abstract}


\maketitle
\section{Introduction}
Phase coherence is one of the fundamental properties of
Bose-Einstein condensates. It is also a key feature in the present developments
of the research on condensates which, ten years after the first experimental realization,
go in the direction of integrating this powerful tool into other branches of physics,
of which metrology and quantum information are two promising examples
\cite{metrology_quantuminfo}.

The problem of the condensate phase dynamics due to atomic interactions
at zero temperature has been analyzed by different authors in theory \cite{PhaseT=0}
and in experiment \cite{MIT,JILA,other_phase_expts}. 
It is now well understood that an initially prepared relative phase between two 
condensates will spread in time due to the corresponding uncertainty 
in the relative particle number as the relative 
phase and the relative particle number are conjugate variables. 
The phase dynamics of a two component condensate in realistic situations
including harmonic traps, non stationarity and fluctuations in the total number 
of particles was analyzed in \cite{mixtures}, where a comparison to the experiments 
of \cite{JILA} is also performed.
An important conclusion was that the zero temperature theory could not account for the 
coherence times observed in experiment, which raises the question
of the role of the non-condensed fraction.

In this paper we address the fundamental problem of phase spreading
of a Bose-Einstein condensate in a finite temperature atomic gas.
In order to obtain simple and general results,
we consider the ideal case of a spatially uniform condensate at thermodynamic 
equilibrium, and we assume that one has access to the first order 
temporal correlation function $\langle a_0^\dagger(t) a_0\rangle$
of the component $a_0$ of the atomic field in the condensate mode.
In real life, the situation is more complex: the atoms are trapped
in harmonic potentials, and the measurement of phase coherence
is a delicate procedure, usually relying on the interference
between two condensates \cite{JILA}.
In the literature two well distinct predictions exist for the long time
spreading of the condensate phase at finite temperature, either a diffusive
behavior (variance growing linearly in time)
\cite{Zoller,GrahamPRL,GrahamPRA,GrahamJMO} or a ballistic behavior
(variance growing quadratically in time) \cite{Kuklov}.
We study this problem first with a classical field model 
\cite{Kagan,Sachdev,Drummond,Rzazewski0,Burnett,Wigner},
where exact numerical simulations can be performed. 
We then explain the numerics 
analytically, and extend the analytical approach to the quantum case.

The important result that we obtain is that the variance of the phase increases
quadratically in time. This is at variance with
the prediction of phase diffusion
from the ``quantum optics" open system approaches of 
\cite{Zoller,GrahamPRL,GrahamPRA,GrahamJMO} assuming the condensate 
to evolve under the influence of Langevin short memory
fluctuating forces.
Our prediction results from two ingredients, (i) 
the system is prepared in an initial state with an energy fluctuating 
from one experimental realization to the other, here sampling the 
canonical ensemble, and (ii) the system is isolated in its further evolution
and therefore keeps a constant energy.
As we shall see, the combination of these two ingredients
prevents some temporal correlation functions to vanish at long times.
Our prediction qualitatively agrees with the one of \cite{Kuklov},
but not quantitatively, as we obtain a different expression for
the long time limit of the variance of the phase over the time squared.
This difference is due to the fact that we take into account 
ergodicity in the system resulting from the interactions among 
Bogoliubov modes such as the Beliaev-Landau processes.

In section \ref{sec:tcfm} we present the classical field model; numerical
predictions for this model are presented in section \ref{sec:cfnr},
and analytical results reproducing the numerics at short or long times
are given in section \ref{sec:cfar}. These analytical results are extended
to the case of the quantum field in section \ref{sec:qtar}. We
conclude in section \ref{sec:c}.

\section{The Classical Field Model}
\label{sec:tcfm}

In this section we develop a classical field model that has the advantage that it can
be exactly simulated numerically. This will allow us to understand the physics
governing the spreading of the condensate phase and to test the validity 
of various approximations, paving the way to the quantum treatment.

We consider a lattice model for a classical field $\psi({\bf r})$ in
three dimensions. The lattice spacings are $l_1$, $l_2$, $l_3$ 
along the three directions 
of space and $dV=l_1 l_2 l_3$ is the volume of the unit cell in the lattice.
We enclose the atomic field in a spatial box of sizes $L_1$, $L_2$, $L_3$
and volume $V=L_1 L_2 L_3$, with periodic boundary conditions.
The discretized field has the following Poisson brackets
\begin{equation}
i\hbar \{\psi({\bf r_1}),\psi^*({\bf r_2})\}=
	\frac{\delta_{{\bf r_1},{\bf r_2}}}{dV} 
\end{equation}
where the Poisson brackets are such that $df/dt = \{f,H\}$ for a
time-independent functional $f$ of the field $\psi$.
The field $\psi$ may be expanded over the plane waves
\begin{equation}
\psi({\bf r})=\sum_{\bf k} a_{\bf k} 
\frac{e^{ \, i \, {\bf k}\cdot{\bf r}}}{\sqrt{V}} \,,
\label{eq:planewaves}
\end{equation}
where $\mathbf{k}$ is restricted to the first Brillouin zone,
$k_\alpha\in [-\pi/l_\alpha,\pi/l_\alpha[$ where $\alpha$ labels the directions
of space.

We assume that, in the real physical system, the total number of atoms is fixed, equal to $N$.
In the classical field model, this fixes the norm squared of the field:
\begin{equation}
dV \sum_{\mathbf{r}} |\psi(\mathbf{r})|^2 = N.
\end{equation}
Equivalently the density of the system 
\begin{equation}
\rho=\frac{N}{V}
\end{equation}
is fixed for each realization of the field.
The evolution of the field is governed by the Hamiltonian
\begin{equation}
H=\sum_{\bf k} \tilde{E}_k {a}_{\bf k}^* a_{\bf k}
+ \frac{g}{2} \sum_{{\bf r}} dV
\psi^*({\bf r})\psi^*({\bf r})
\psi({\bf r}) \psi({\bf r}),
\label{eq:Hamiltonian}
\end{equation}
where $\tilde{E}_k$ is the dispersion relation of the non-interacting waves,
and the binary interaction between particles in the real gas is reflected in the classical
field model by a field self-interaction with a coupling constant
$g=4\pi \hbar^2 a/m$, where $a$ is the $s$-wave scattering length of
two atoms.

In general, we expect the predictions of a classical field model to be cut-off
dependent, i.e.\ the predictions of our model may depend on the lattice 
spacings $l_\alpha$.
We use here a refinement to
the usual classical field model, which makes it cut-off independent for
{\it some} observables like the condensate fraction, a quantity expected to play
an important role here. An obvious example of a quantity which will remain cut-off dependent
is the mean value of the Hamiltonian $H$ in thermal equilibrium.

Let us consider first the non-interacting case ($g=0$) in presence of a condensate.
For a thermalized classical field
the occupation numbers of the excited plane wave modes are given by the equipartition formula
\begin{equation}
\langle a_k^* a_k \rangle
= \frac{k_B T}{\tilde{E}_k}\, .
\end{equation}
We adjust the dispersion relation $\tilde{E}_k$ in order to reproduce
the Bose law for the occupation numbers of the quantum field in the Bose-condensed
regime:
\begin{equation}
\frac{1}{e^{\beta \hbar^2 k^2/2m} -1} = \frac{k_B T}{\tilde{E}_k} \,.
\label{eq:Ektilde}
\end{equation}
For all modes with large occupation number $\tilde{E_k} \simeq \hbar^2 k^2/2m$,
while the occupation of modes with $\hbar^2 k^2/2m \gg k_B T$, whose quantum dynamics is
not well approximated by the classical field model anyway, 
is exponentially suppressed as in the quantum theory.  

In the interacting case, one could adapt the same trick of a modified dispersion relation,
by including the fact that the relevant spectrum is not $\hbar^2 k^2/2m$ but
the Bogoliubov spectrum \cite{trick}.
The resulting $\tilde{E}_k$ would now start growing exponentially
with $k$ when the Bogoliubov energy $[(\hbar^2 k^2/2m)(2\rho g+\hbar^2 k^2/2m)]^{1/2}$
reaches $k_B T$.

In the classical field model
we restrict our analysis
to the regime $k_B T \gg \rho g$ so that at energies of the order of $k_B T$,
the Bogoliubov energy is dominated by the kinetic term $\hbar^2 k^2/2m$.
One can then simply use in the Hamiltonian the modified dispersion 
relation $\tilde{E}_k$ as given by Eq.(\ref{eq:Ektilde}).
This is what we did in the simulations of this paper, so that the classical field $\psi$
evolves according to the non-linear equation \cite{Delta}:
\begin{equation}
i \hbar \, \partial_t \psi= \left\{k_B T
\left[\exp\left(-\beta\frac{\hbar^2}{2m}\Delta\right)-1
\right]
+ g |\psi({\bf r},t)|^2  \right\} \, \psi\, .
\label{eq:sc}
\end{equation}
In practice this equation is integrated numerically with the FFT splitting technique.

We then introduce the density and the phase of the condensate mode
\begin{equation}
a_0=e^{\, i \, \theta} \sqrt{N_0}\,.
\label{eq:intphase}
\end{equation}
In what follows, we concentrate on three physical quantities:
the condensate amplitude correlation function
\begin{equation}
\langle a_0^*(t) \, a_0(0) \rangle \,,
\label{eq:corra0}
\end{equation}
the condensate atom number correlation function
\begin{equation}
\langle \delta N_0(t) \, \delta N_0(0) \rangle
\hspace{0.5cm} \mbox{where} \hspace{0.5cm}
\delta N_0=N_0 - \langle N_0 \rangle \,,
\label{eq:corrn0}
\end{equation}
and the variance of the condensate phase change during $t$:
\begin{equation}
\mbox{Var}\, \varphi(t)=\langle {{\varphi}}(t)^2 \, \rangle -
\langle {{\varphi}}(t) \, \rangle^2
\hspace{0.5cm} \mbox{where} \hspace{0.5cm}
{{\varphi}}(t)={\theta}(t)-{\theta}(0).
\label{eq:varphase}
\end{equation}
The averages are taken over stochastic realizations
of the classical field, as the 
initial field samples a thermal probability distribution.

\section{Classical field: Numerical Results}
\label{sec:cfnr}

We consider a gas of $N=4\times10^5$ atoms with $\rho g=700 \, \hbar^2/m V^{2/3}$ 
in a box of non 
commensurable square lengths to guarantee efficient ergodicity in the system, in the ratio
$L_1^2 : L_2^2 : L_3^2 =  \sqrt{2} : (1+\sqrt{5})/2 : \sqrt{3}$.
We choose the number of the lattice points in a temperature dependent way,
such that the maximal Bogoliubov 
energy $[(\hbar^2 k^2/2m)(2\rho g+\hbar^2k^2/2m)]^{1/2}$ on the lattice is equal to $3 k_BT$.

To generate the stochastic initial values of the classical field
we proceed as follows. 
(i) For each realization, 
we generate a non condensed field $\psi_\perp({\bf r})$ at temperature $T$
in the Bogoliubov approximation as explained in \cite{Cartago}. 
In practice we generate complex numbers $\{ b_{\bf k} \}$ for each vector $\bf k$
on the grid according to the probability distribution
\begin{equation}
P(b_{\bf k})= \frac{1}{\pi} \frac{\tilde{\epsilon}_{k}}{k_BT} \, 
	e^{- (|b_{\bf k}|^2 \, \tilde{\epsilon}_{k}/k_BT) }
\end{equation}
where  $ \tilde{\epsilon}_{k}=[\tilde{E}_{k}(\tilde{E}_{k}+2\rho g)]^{1/2}.$
With a set of $\{ b_{\bf k} \}$ for a given realization 
we build the non condensed field
\begin{equation}
\psi_\perp({\bf r})= e^{i\theta} \sum_{\mathbf{k}\neq\mathbf{0}} \left(b_{\bf k} \tilde{U}_k \frac{e^{i {\bf k} \cdot {\bf r}}}{\sqrt{V}}
 + b_{\bf k}^\ast \tilde{V}_k \frac{e^{-i {\bf k} \cdot {\bf r}}}{\sqrt{V}}\right)
\label{eq:decomp}
\end{equation}
where the initial value of the condensate phase $\theta$
is randomly chosen with the uniform law
in $[0,2\pi[$, and 
where the real amplitudes $\tilde{U}_k$, $\tilde{V}_k$, normalized
as $\tilde{U}_k^2-\tilde{V}_k^2=1$, are given by the usual Bogoliubov theory, here with 
the modified dispersion relation, so that 
\begin{equation}
\tilde{U}_k+\tilde{V}_k = \left(\frac{\tilde{E}_k}{\tilde{E}_k+2\rho g}\right)^{1/4}\ .
\end{equation}
(ii) We create the classical field with the constraint that the total number of atoms $N$ is fixed:
\begin{equation} 
\psi({\bf r})=\frac{a_0}{\sqrt{V}}+\psi_\perp({\bf r})
\label{eq:split}
\end{equation} 
where $a_0=\sqrt{N-N_\perp} e^{i\theta}$, $N_\perp$ is the number of non condensed atoms,
\begin{equation}
N_\perp=\sum_{\bf r} dV |\psi_\perp({\bf r})|^2\ .
\label{eq:defnperp}
\end{equation}
(iii) We let the field evolve for some time interval with the Eq.(\ref{eq:sc})
to eliminate transients due to the fact that the Bogoliubov approximation used in the
sampling does not produce an exactly stationary distribution.
After this `thermalization' period we start calculating the relevant observables,
as $\psi$ evolves with the same Eq.(\ref{eq:sc}).

First we investigate the mean condensate phase
change $\langle \varphi\rangle(t)$. We find a linear dependence with time,
with a slope slightly different from the value
$-\rho g/\hbar$ naively expected, e.g.\ from the zero temperature Gross-Pitaevskii equation. 
The slope difference is temperature dependent and is expected physically to correspond
to the discrepancy between the zero temperature chemical potential $\rho g$
and the actual finite temperature one $\mu(T)$. This we shall confirm using Bogoliubov
theory in Sec.~\ref{sec:cfar} (see also \cite{Rzazewski2}).

\begin{figure}[htb]
\centerline{\includegraphics[width=7cm,clip=]{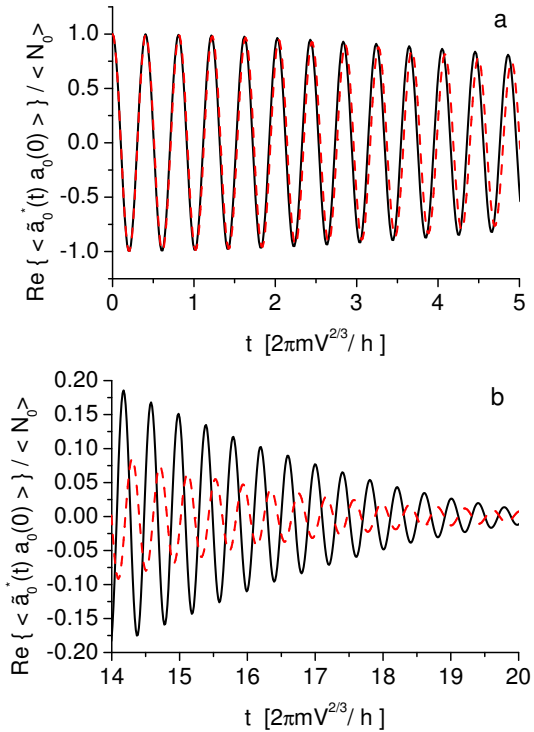}}
\caption{(Color online) Real part of the condensate amplitude correlation function (\ref{eq:corra0}) normalized
to its $t=0$ value, and divided
by the zero temperature evolution $e^{i\rho g t/\hbar}$,
as a function of time: in the vertical axis label, $\tilde{a}_0(t)$ stands for
$a_0(t) e^{i\rho g t/\hbar}$.
(a) Short times behavior and (b) long times behavior.
In solid line from an average over 500 solutions of Eq.(\ref{eq:sc}),
in dashed line (red)
the Bogoliubov approximation (\ref{eq:Bogcorra0}).
Here the temperature is $k_B T=3077.3 \, \hbar^2/m V^{2/3}=0.1711 T_c$, 
where $T_c$ is the critical 
temperature $k_B T_c= (2\pi \hbar^2/m)(\rho/\zeta(3/2))^{2/3}$ of the ideal gas,
the number of particles is $N=4\times 10^5$ and the coupling constant is such
that the zero-temperature chemical potential is $\rho g=g N/V=700 \hbar^2/m V^{2/3}$.
}
\label{fig:16a0}
\end{figure}

In figure \ref{fig:16a0}, we show the real part
of the amplitude correlation function of the condensate 
$\langle a_0^*(t) a_0(0) \rangle$ as a function of time, for a
temperature $T=0.17 T_c$, where $T_c$ is the critical temperature 
of the ideal gas.
The zero-temperature evolution $e^{i\rho g t/\hbar}$ is removed
so that the oscillations in the figure are due to the above mentioned
effect $\mu(T)\neq \rho g$.
Due to the finite temperature in the system,
the correlation function of the condensate amplitude is smeared 
out at long times. 

\begin{figure}[htb]
\centerline{\includegraphics[width=8cm,clip=]{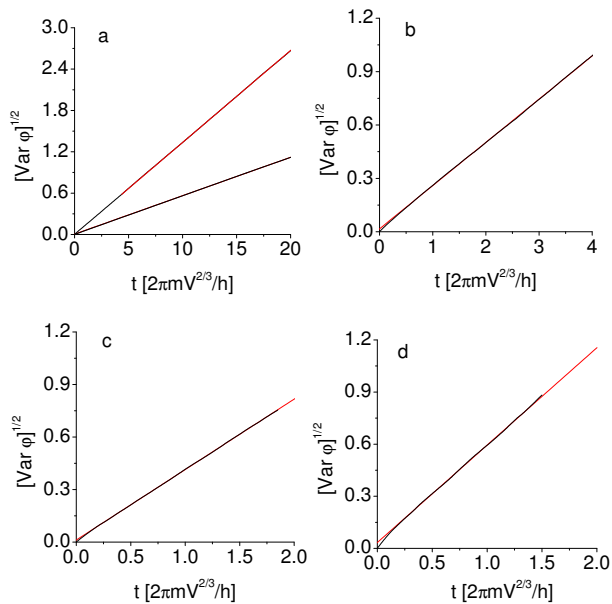}}
\caption{(Color online) Standard deviation of the condensate phase change  $\varphi(t)$
(\ref{eq:varphase}) as a function of time
for (a) $T=0.08245 T_c$ (lower curve) and $T=0.1711 T_c$ (upper curve), (b) $T=0.29467 T_c$,
(c) $T=0.453 T_c$, (d) $T=0.6473 T_c$. 
Thick solid line (black): 
numerical solution from the classical field model Eq.(\ref{eq:sc}) averaged over 500 realizations.
Thin solid line (red): a linear fit.
The parameters $N$ and $\rho g$ have the same values as in Fig.\ref{fig:16a0}.}
\label{fig:16etoth_var_th}
\end{figure}

Correspondingly the standard deviation of the condensate phase change increases with time,
as we show in figure \ref{fig:16etoth_var_th} for five different values
of the temperature, up to $T=0.65 T_c$.
In all cases, at long times,
we observe a quadratic growth of $\mbox{Var}\, \varphi$ contrarily to the
phase diffusion behavior $\propto t$ predicted in the literature \cite{Zoller,GrahamPRL,GrahamPRA,GrahamJMO}.

\begin{figure}[htb]
\centerline{\includegraphics[width=7cm,clip=]{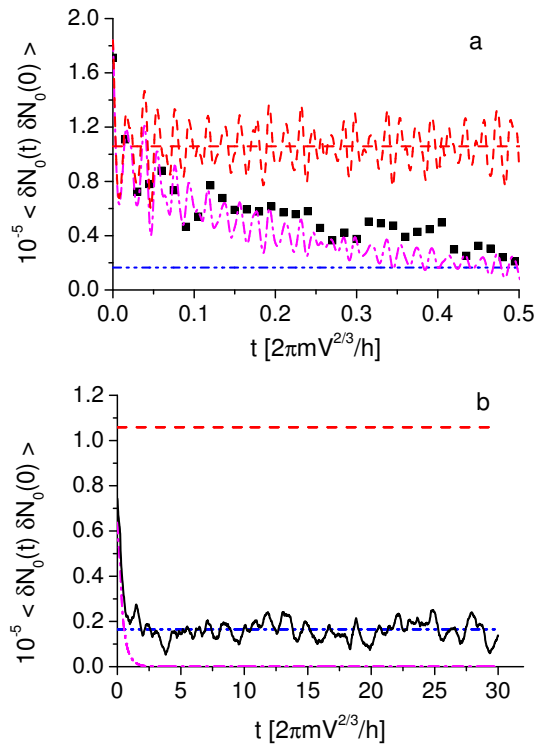}}
\caption{(Color online) Correlation function of the condensate atom number (\ref{eq:corrn0}).
(a) Short times, (b) long times. 
The classical field results are obtained from an average over 500 solutions of Eq.(\ref{eq:sc}); 
they are represented by symbols in (a) and a solid line in (b).
The dashed lines (red) are the Bogoliubov approximations (\ref{eq:Bogcorrn0})
(oscillating line) and (\ref{eq:Bogcorrn0_nonosc}) (horizontal line) 
in (a), and only (\ref{eq:Bogcorrn0_nonosc}) in (b). 
The dashed-dotted line (purple) is the Gaussian model.
For clarity in (b) we washed out fast oscillations 
in the simulation result and in the Gaussian model,
by averaging over consecutive points over
a time width $0.45 m V^{2/3}/\hbar$.
The horizontal dashed-dotted-dotted line (blue) in (a) and (b) is
the ergodic long time limit prediction, described in section
\ref{sec:cfar}.
The parameters $N$, $T$ and $\rho g$ have the same values as in Fig.\ref{fig:16a0}.}
\label{fig:16n0}
\end{figure}

To complete the physical picture,  we show in figure \ref{fig:16n0} the 
correlation function of the condensate atom number (\ref{eq:corrn0}). 
At very short times, see the beginning of the curves in Fig.\ref{fig:16n0}a, 
the simulation (square symbols) confirms the Bogoliubov prediction
(dashed oscillating line); at long times, see Fig.\ref{fig:16n0}b,
the correlation function drops to a value significantly smaller
than the Bogoliubov prediction (fast oscillations are not shown in the figure);
a key point is that this long time value of the correlation function
of the condensate atom number is not zero.

\begin{figure}[htb]
\centerline{\includegraphics[width=7cm,clip=]{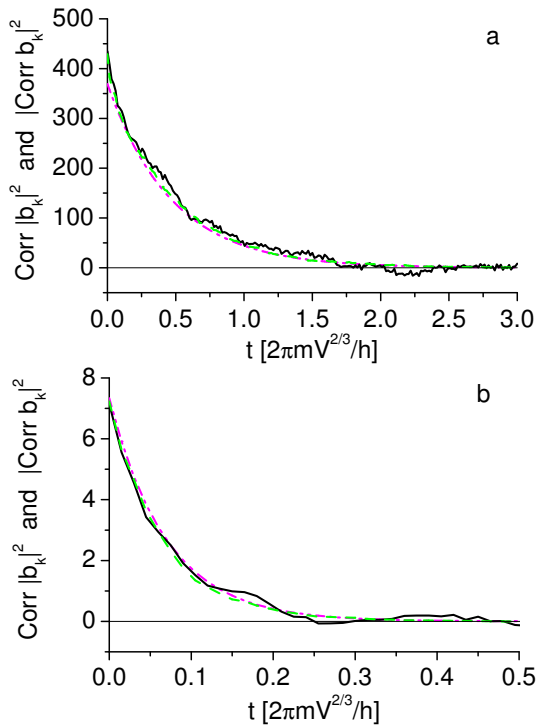}}
\caption{(Color online) 
Correlation function of $|b_\mathbf{k}|^2$ (black solid line)
and modulus squared of the correlation function of $b_\mathbf{k}$
(green dashed line),
as obtained from 2000 realisations of the classical field simulations,
(a) for the mode ${\bf k}=(0, 0, 2\pi/L_3)$
and (b) for the mode ${\bf k}=(2\pi/L_1, 10\pi/L_2, 4\pi/L_3)$. 
We define the correlation function of a quantity $X$
as $\mbox{Corr}\, X= \langle X^*(t)X(0)\rangle - |\langle X\rangle|^2$.
The purple dashed-dotted line is an exponential function of $t$,
given by Eq.(\ref{eq:ffg}).
It reproduces well the simulation results,
and we have checked that the agreement is good
for all vectors ${\bf k}$ lying in an arbitrarily chosen plane in $k$-space.
The parameters $N$, $T$ and $\rho g$ have the same values as in Fig.\ref{fig:16a0}.}
\label{fig:24bk}
\end{figure}

One may fear at this stage that the classical field model is 
missing some source of damping in the dynamics of the system.
However it is a well established fact that the classical field
model is able to
simulate damping processes, including the finite temperature 
Beliaev-Landau processes \cite{Vincent,Stringari_Pitaevskii,Giorgini,Shlyapnikov}, 
since the interaction among the Bogoliubov
modes is included in this model
\cite{Kagan,Sachdev,DaliShlyap,Rzazewski0,Burnett,Cartago,Rzazewski1,Rzazewski2,
vortex_formation,bec_collision}.
More quantitatively
we now check that the damping times due to the Beliaev-Landau
processes in the simulation are much shorter than
the evolution times considered here. 
To this end, we extract from the simulations the temporal
correlation functions $\langle b_\mathbf{k}^*(t) b_\mathbf{k}(0)\rangle$
and $
\langle |b_\mathbf{k}|^2(t) |b_\mathbf{k}|^2(0)\rangle
-\langle |b_\mathbf{k}|^2\rangle^2$,
obtained by projecting the classical field over the corresponding
Bogoliubov mode and averaging over many realizations.
We show these correlation functions for the lowest energy
Bogoliubov mode and for an excited Bogoliubov mode in Fig.\ref{fig:24bk}.

We come then into a paradox. On one side, the various Bogoliubov
oscillators $b_\mathbf{k}$ decorrelate at long times. On the other
side, the variance of the phase change $\varphi$ of the condensate
varies quadratically at long times, which implies, as we shall see
in Sec.~\ref{sec:cfar}, that the derivative of the phase $\dot\varphi$
does not decorrelate at long times, although it is a function
of the $b_\mathbf{k}$'s; similarly, the fluctuations of the number
of condensate atoms $\delta N_0$, which are functions of the $b_\mathbf{k}$'s,
do not decorrelate at long times.

\begin{figure}[htb]
\centerline{\includegraphics[width=7cm,clip=]{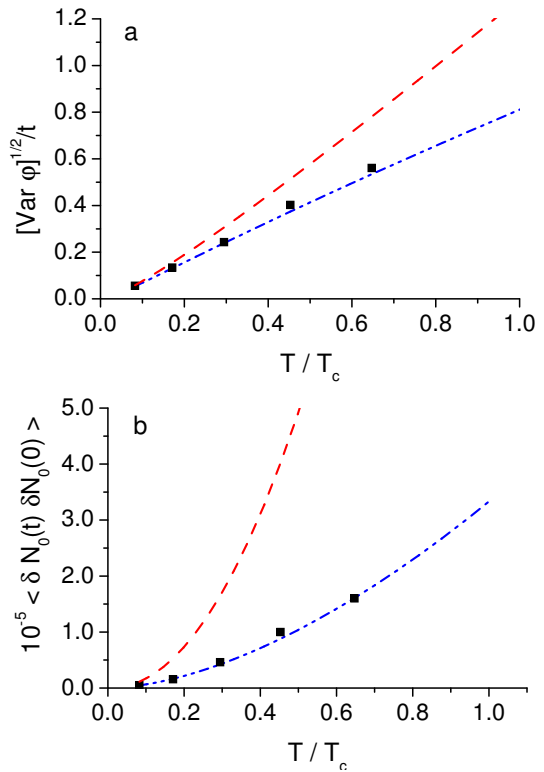}}
\caption{(Color online) (a)  Long time limit of
$(\mbox{Var}\,\varphi)^{1/2}/t$, in units of $\hbar/m V^{2/3}$,
as a function of the temperature over the ideal gas critical temperature.
(b) Long time limit of the condensate atom number correlation function,
as a function of $T/T_c$.
In square symbols the numerical results of the classical field simulation
(averaged over 500 realizations), 
in dashed line (red) the Bogoliubov prediction and in dashed-dotted-dotted (blue)
the ergodic prediction. 
The parameters $N$ and $\rho g$ have the same values as in Fig.\ref{fig:16a0}.
}
\label{fig:tphi}
\end{figure}

This paradox will be explained in Sec.~\ref{sec:cfar},
and quantitative predictions for 
long times behavior of the condensate 
atom number correlation function and of the 
variance of the condensate phase change will be derived.
Anticipating these analytical results,  we show in Fig.\ref{fig:tphi}a
the long time limit of $(\mbox{Var}\, \varphi)^{1/2}/t$
as a function of $T/T_c$, from
the results of the classical field simulations,
but also from the predictions of the Bogoliubov
approximation Eq.(\ref{eq:Bogvarphase}), and of the ergodic theory
of Sec.~\ref{sec:cfar}.
In figure  \ref{fig:tphi}b we show the same results and predictions
for the asymptotic value 
of the condensate atom number correlation function.

\section{Classical field: Analytical results}
\label{sec:cfar}

The general procedure used here to obtain analytical results
is the following. First one expresses the quantity of interest (the number of
condensate atoms or the time derivative of the condensate
phase) in terms on the amplitudes $b_\mathbf{k}$ of the field $\psi$ over the Bogoliubov
modes,
\begin{eqnarray}
b_{\mathbf{k}}(t) &=& dV\sum_\mathbf{r} 
\tilde{U}_k \frac{e^{-i\mathbf{k}\cdot\mathbf{r}}}{\sqrt{V}} e^{-i\theta(t)} \psi_\perp(\mathbf{r},t) 
\nonumber \\
&& +
\tilde{V}_k \frac{e^{i\mathbf{k}\cdot\mathbf{r}}}{\sqrt{V}} e^{i\theta(t)} \psi_\perp^*(\mathbf{r},t) 
\end{eqnarray}
where $\psi_\perp$ is the component of $\psi$ orthogonal to the condensate mode.
Second one evaluates the correlation functions of products of
$b_\mathbf{k}$ in various physical limits.

\subsection{Correlation function of the condensate atom number}
\label{subsec:corr_n0}

As the total number of particles is fixed, it is equivalent to
calculate the correlation function of $\delta N_0$ in Eq.(\ref{eq:corrn0})
and of the number of non-condensed particles $N_\perp$.
Injecting the expansion Eq.(\ref{eq:decomp}) for the time dependent non-condensed
field $\psi_\perp$ over the Bogoliubov modes into Eq.(\ref{eq:defnperp}) we obtain
\begin{equation}
\label{eq:nperp}
N_\perp(t) = \sum_{\mathbf{k}\neq\mathbf{0}} |\tilde{U}_k b_{\bf k}(t) + \tilde{V}_k b_{\bf -k}^*(t)|^2.
\end{equation}

\noindent{\it Bogoliubov theory:} 
In Bogoliubov theory interaction among the Bogoliubov modes is neglected so that
at all times 
\begin{equation}
b_{\bf k}(t)=b_{\bf k}(0)\, e^{-i \omega_k t} \hspace{1cm} \mbox{with}
\hspace{1cm} \omega_k=\tilde{\epsilon_k}/\hbar \,.
\label{eq:bBogol}
\end{equation}
As Wick's theorem applies for the initial thermal distribution we obtain
for the correlation function of the condensate atom number:
\begin{eqnarray}
\langle \delta N_0(t) \, \delta N_0(0) \rangle_{\rm Bog}  &=&
\sum_{\bf k\neq\bf 0} \tilde{n}_k^2 \left[ |\tilde{U}_k^2 e^{i\omega_k t} + \tilde{V}_k^2 e^{-i\omega_kt}|^2 \right.\nonumber \\
&&\left. + (\tilde{U}_k \tilde{V}_k)^2 |e^{i\omega_k t} + e^{-i\omega_kt}|^2 \right] 
\label{eq:Bogcorrn0}
\end{eqnarray}
where $\tilde{n}_k=k_B T/\tilde{\epsilon}_k$ is the Bogoliubov mean occupation number of a mode
for the classical field.
At very short times, a good agreement of the Bogoliubov prediction with the
simulation is observed in Fig.\ref{fig:16n0}a.
Smearing out the terms oscillating rapidly at Bohr frequencies $2\omega_k$, 
we obtain a prediction directly comparable to the coarse grained numerical
result of Fig.\ref{fig:16n0}b: 
\begin{equation}
\langle \delta N_0(t) \, \delta N_0(0) \rangle_{\rm Bog \; non \, osc} =
\sum_{\bf k{\neq} 0} (\tilde{U}_k^2 + \tilde{V}_k^2)^2 \, \tilde{n}_k^2 \,.
\label{eq:Bogcorrn0_nonosc}
\end{equation}
This amounts to considering the correlation function of 
\begin{equation}
N_\perp^{\rm non\ osc}(t) = \sum_{\mathbf{k}\neq\mathbf{0}}
(\tilde{U}_k^2+\tilde{V}_k^2) b_\mathbf{k}^*(t) b_\mathbf{k}(t),
\end{equation}
deduced from (\ref{eq:nperp}) by eliminating the oscillating terms
such as $b_\mathbf{k} b_{-\mathbf{k}}$.
As can be seen in Fig.\ref{fig:16n0}b, Bogoliubov theory fails at long times.
Note that in the thermodynamic limit, where the above sum is dominated
by the low $k$ terms, one may approximate $\tilde{V}_k\sim -\tilde{U}_k$, so that 
Eq.(\ref{eq:Bogcorrn0_nonosc}) is roughly half of the $t=0$ 
value of Eq.(\ref{eq:Bogcorrn0}); in other words, it is approximately half of the
variance of the condensate number. In the numerical result of Fig.\ref{fig:16n0},
the correlation function drops by much more than a factor $2$.

\noindent{\it Gaussian theory:} 
A possible approach to improve Bogoliubov theory consists in assuming
that the $b_\mathbf{k}$ are Gaussian variables with a finite time correlation
due to the Beliaev-Landau mechanism:
\begin{equation} 
|\langle b_k^\ast (t) b_k(0) \rangle|^2=\tilde{n}_k^2 \, e^{-2 \gamma_k |t|}
\label{eq:ffg}
\end{equation} 
where $\gamma_k$ is calculated with time dependent perturbation theory including
the discrete nature of the spectrum as in
\cite{Cartago}. This amounts to weighting each term of Eq.(\ref{eq:Bogcorrn0_nonosc})
by $\exp(-2\gamma_k |t|)$. 
This assumption is supported by numerical evidence for a single mode, see Fig.\ref{fig:24bk},
and by an analytic derivation in the thermodynamic limit for one or two modes, 
see Appendix \ref{appen:pilote}.
Nevertheless, the resulting prediction for the correlation function of $N_0$, 
while looking promising at
short times, see Fig.\ref{fig:16n0}a, is in clear disagreement
with the simulation at long times, see Fig.\ref{fig:16n0}b. 
Since the assumption of a long time
decorrelation of $b_\mathbf{k}^*(t)$ with $b_\mathbf{k}(0)$ is physically reasonable,
one may suspect that the Gaussian hypothesis is not accurate when a large
number of modes are involved as for the correlation function of $N_0$. 
This is indeed the case, as we now show.

\noindent{\it Ergodic theory:} 
A systematic way to calculate the long time limit of the correlation function
is to assume that the non-linear dynamics generated by Eq.(\ref{eq:sc})
is ergodic: at long times, the $b_\mathbf{k}(t)$'s for a given realization of the 
field explore uniformly a fixed energy surface in phase space 
\cite{moment}. In the Bogoliubov 
approximation for the energy, this means that the $b_\mathbf{k}(t)$'s
sample the unnormalized probability distribution
\begin{equation}
P_\infty(\{b_\mathbf{k}\}) = \delta\left(E-\sum_{\mathbf{k}\neq\mathbf{0}}
\tilde{\epsilon}_k b_\mathbf{k}^* b_\mathbf{k}\right)
\label{eq:ergo}
\end{equation}
where the Bogoliubov energy $E$ is fixed by the initial value of the field:
\begin{equation}
E = \sum_{\mathbf{k}\neq\mathbf{0}} \tilde{\epsilon}_k b_\mathbf{k}^*(0) b_\mathbf{k}(0).
\label{eq:E=}
\end{equation}

First, for a given initial condition of the field, we calculate the expectation
value of $N_\perp(t)$ as given by Eq.(\ref{eq:nperp})
over the ergodic distribution Eq.(\ref{eq:ergo}), which is equivalent to
the temporal average of $N_\perp(t)$ over an infinite time interval. 
The terms of the form $b\, b$
or $b^*\, b^*$ have a zero mean, since the phases of the $b_\mathbf{k}$'s are uniformly
distributed over $2\pi$, according to Eq.(\ref{eq:ergo}).
To calculate the expectation value of the $b^* b$ terms, it is convenient to introduce 
rescaled variables 
\begin{equation}
B_\mathbf{k}= \left(\frac{\tilde{\epsilon}_k}{E}\right)^{1/2} \, b_\mathbf{k}.
\end{equation}
According to Eq.(\ref{eq:ergo}) the real parts and the imaginary parts of all the
$B_\mathbf{k}$ are uniformly distributed over the unit hypersphere in a space of
dimension $2\mathcal{M}$, where $\mathcal{M}=V/dV-1$ is the number of Bogoliubov modes
so that we obtain $ \overline{|B_\mathbf{k}|^2}
=1/{\mathcal{M}}$ where the overline stands for the average over the ergodic distribution
(\ref{eq:ergo}). As a consequence the ergodic average of $N_\perp$ is
\begin{equation}
\overline{N_\perp}=
\frac{1}{\mathcal{M}}
\sum_{\mathbf{k}\neq\mathbf{0}} (\tilde{U}_k^2+\tilde{V}_k^2) \frac{E}{\tilde{\epsilon}_k}.
\end{equation}
Note that this ergodic average depends on the $t=0$ value of the $b_\mathbf{k}$'s
{\it via} (\ref{eq:E=}).

Second, we average the product $\overline{N_\perp}\, N_\perp(0)$ over the thermal
canonical distribution for the initial values $b_\mathbf{k}(0)$.
This gives the long time limit of the correlation function of the number of condensate
atoms:
\begin{equation}
\langle \delta N_0(t\rightarrow +\infty)\delta N_0(0)\rangle_{\rm ergo} =
\frac{1}{\mathcal{M}} \left(\sum_{\mathbf{k}\neq \mathbf{0}}
(\tilde{U}_k^2+\tilde{V}_k^2) \tilde{n}_k\right)^2.
\label{eq:dn0c_ergo}
\end{equation}
This prediction is in good agreement with the simulations at long times,
see Fig.\ref{fig:16n0}b for a fixed value of the temperature,
and Fig.\ref{fig:tphi}b as a function of temperature.
Note that, according to Schwartz inequality, the ergodic value is lower than
the coarse grained Bogoliubov prediction Eq.(\ref{eq:Bogcorrn0_nonosc}),
as was expected physically.

This clearly shows that the existence of infinite time correlations in the
number of condensate atoms is a consequence of the conservation of energy
during the free evolution of the system.

To understand the failure of the Gaussian model, 
we give the ergodic prediction of the long-time
limit of the correlation function of the Bogoliubov mode occupation numbers
$n_{\bf k}=|b_{\bf k}|^2$,
\begin{equation}
\langle n_\mathbf{k}(t\to+\infty) n_{\mathbf{k}'}(0)\rangle_{\rm ergo}
-\langle  n_\mathbf{k} \rangle \langle n_{\mathbf{k}'} \rangle =
\frac{\tilde{n}_k \tilde{n}_{k'}}{\mathcal{M}}.
\label{eq:dndnergo}
\end{equation}
This long-time value is non-zero, contrarily to the Gaussian model prediction.
One may argue that the value Eq.(\ref{eq:dndnergo}) tends to zero in the thermodynamic limit, 
so that the error in the Gaussian model looks negligible for a large system.
However, in calculating the correlation function of a macroscopic quantity such
as $N_\perp$, a {\it double} sum over the Bogoliubov modes appears, so that 
the small deviations Eq.(\ref{eq:dndnergo}) from the Gaussian model prediction
sum up to a macroscopic value.
In other words, in the calculation of a given correlation function, one is not allowed
to take the thermodynamic limit before the end of the calculation.

\subsection{Variance of the condensate phase change}

To reproduce the approach of the previous subsection for the phase, one should express
the phase change $\varphi(t)$ of the condensate amplitude $a_0$ as a function
of the $b_\mathbf{k}$'s. It turns out that the quantity easily expressed
in terms of the $b_\mathbf{k}$'s is the time derivative $\dot{\varphi}$.
The variance of $\varphi$ is then related to the correlation function $\mathcal{C}$
of $\dot\varphi$:
\begin{equation}
\mbox{Var}\, {\varphi}
 = \int_0^t \, d\tau \, \int_0^t \, d\tau^\prime\,
\mathcal{C}(|\tau-\tau^\prime|)
\end{equation}
where time translational invariance in steady state
imposes for a classical field that ${\cal C}$ depends only on $|\tau-\tau^\prime|$:
\begin{equation}
{\cal C}(|\tau-\tau^\prime|)  =
\langle \dot{\varphi}(\tau) \dot{\varphi}(\tau^\prime) \rangle -
       \langle \dot{\varphi}(\tau) \rangle 
       \langle \dot{\varphi}(\tau^\prime) \rangle \,.
\end{equation}
If $\mathcal{C}(\tau)\rightarrow 0$ fast enough
when $\tau \rightarrow \infty$
then $\mbox{Var}\,{\varphi}$ grows linearly in time.  
On the other hand, if $\mathcal{C}(\tau)$ has a non-zero limit at long times,
then $\mbox{Var}\, {\varphi}$ grows quadratically in time
\cite{fastenought}.

To express $\dot{\varphi}$ in terms of the $b_\mathbf{k}$'s, 
we write the equation of motion for $a_0$:
\begin{eqnarray}
i \hbar \dot{a}_0 &=& i \hbar \{ a_0,H\} = \partial_{a_0^*} H \nonumber \\
&=& \frac{g}{\sqrt{V}} \sum_\mathbf{r} \psi^*(\mathbf{r}) \psi^2(\mathbf{r})
\end{eqnarray}
where we used $\partial_{a_0^*} \psi^*(\mathbf{r})=1/\sqrt{V}$ obtained from
Eq.(\ref{eq:planewaves}).
We split $\psi$ as in Eq.(\ref{eq:split}); we eliminate the condensate amplitude
in the resulting expression for $\dot{a}_0/a_0$ 
(i) by using $|a_0|^2=N-N_\perp$, where $N_\perp$ is a function of the $b_\mathbf{k}$'s,
see Eq.(\ref{eq:nperp}), and (ii) by introducing the field 
\cite{CastinDum}
\begin{equation}
\Lambda({\bf r})=e^{-i \, \theta} \psi_\perp({\bf r}) 
\end{equation}
which is a function of the $b_\mathbf{k}$'s only according to 
Eq.(\ref{eq:decomp}).
This leads to
\begin{eqnarray}
i\hbar \frac{\dot{a_0}}{a_0} &=& \rho g+ \frac{g}{V}\sum_{\bf r} dV \left[ \Lambda({\bf r})^2 +
	|\Lambda({\bf r})|^2  \right] \nonumber \\
&+&
	\frac{g}{\sqrt{V}} \sum_{\bf r} dV 
\frac{\Lambda^*({\bf r}) \Lambda^2({\bf r})}{\sqrt{N-N_\perp}}.
\label{eq:a0d}
\end{eqnarray}
The real part of the above equation gives $-\hbar \dot{\theta}$, which is also
$-\hbar \dot\varphi$.

Restricting to a weak non-condensed fraction, we drop the cubic terms in
Eq.(\ref{eq:a0d}), to obtain \cite{notN0}
\begin{eqnarray}
\hbar\dot{\varphi}  &\simeq & -\rho g-\frac{1}{2}\frac{g}{V} \sum_{\bf r} dV 
	\left[ \Lambda({\bf r})+\Lambda^*({\bf r}) \right]^2 \nonumber \\
&= & -\rho g - \frac{1}{2}\frac{g}{V} \sum_{\mathbf{k}\neq\mathbf{0}}   \,
        (\tilde{U}_k+\tilde{V}_k)^2 \, |b_{\bf k}+b_{\bf -k}^\ast|^2 .
\label{eq:thetadot}
\end{eqnarray}
It turns out that the products 
$b_{\bf k} b_{\bf -k}$ generate oscillating terms 
which do not contribute to a coarse grained time average. 
It is thus useful to define
\begin{equation}
\hbar\dot{\varphi}_{\rm non\ osc}= -\rho g -
\frac{g}{V} \sum_{\mathbf{k}\neq\mathbf{0}}   \, (\tilde{U}_k+\tilde{V}_k)^2 \, |b_{\bf k}|^2.
\label{eq:thetadot_nonosc}
\end{equation}

\noindent{\it Bogoliubov theory}:
By using (\ref{eq:bBogol}) and Wick's theorem we calculate the correlation
function of Eq.(\ref{eq:thetadot}). By temporal integration we
obtain the variance of the condensate phase change
\begin{equation}
(\mbox{Var}\, {\varphi})_{\rm Bog} = 
\left(\frac{g}{\hbar V}\right)^2 \sum_{\mathbf{k}\neq\mathbf{0}}
(\tilde{U}_k+\tilde{V}_k)^4 \, 
\tilde{n}_k^2 \, \left[t^2 + \frac{\sin^2 \omega_k t}{(2 \omega_k)^2} 
\right]\,.
\label{eq:Bogvarphase}
\end{equation}
Qualitatively Bogoliubov theory correctly predicts
a quadratic growth of the variance of $\varphi$
at long times. 
As we show in Fig.\ref{fig:tphi}a, however, it is not fully
quantitative: it does not reproduce the value of the dephasing rate
obtained from the simulations.
This is not surprising as in the full non linear theory 
the $b_{\bf k}$'s interact and do not follow Eq.(\ref{eq:bBogol}).

To be complete, we also give the Bogoliubov approximation for
the correlation function of the condensate amplitude $a_0$.
Neglecting the fluctuations of the modulus of $a_0$, 
one can set
\begin{equation}
\langle a_0^\ast(t) \, a_0(0) \rangle
\simeq \langle N_0 \rangle \langle e^{-i \, \varphi(t)} \rangle \,.
\end{equation}
Dropping the oscillating terms in $b_{\bf k} b_{\bf -k}$ and
$b^*_{\bf k} b^*_{\bf -k}$ in $\dot{\varphi}(t)$,
which give a small contribution, we get
\begin{equation}
\langle a_0^\ast(t) \, \hat{a}_0(0) \rangle_{\rm Bog}
\simeq \langle N_0\rangle
\prod_{\mathbf{k}\neq\mathbf{0}} \frac{1}{1+ i \frac{g}{\hbar V}(\tilde{U}_k+\tilde{V}_k)^2 \, \tilde{n}_k \, t} \,.
\label{eq:Bogcorra0}
\end{equation}
The resulting expression is plotted as a dashed line in Fig.\ref{fig:16a0}
against the result of the simulation.

\begin{figure}[htb]
\centerline{\includegraphics[width=7cm,clip=]{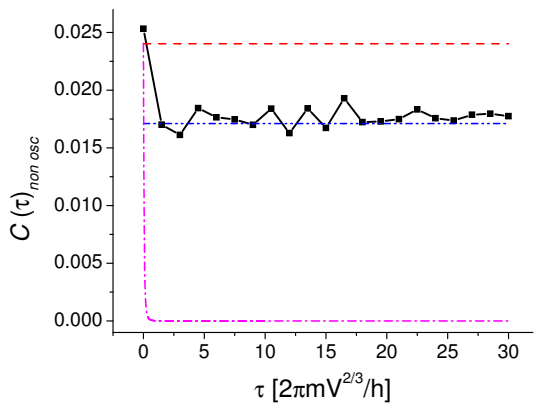}}
\caption{(Color online) Correlation function ${\cal C}_{\rm non\ osc}$ of
the quantity $\dot\varphi_{\rm non\ osc}$ defined in 
Eq.(\ref{eq:thetadot_nonosc})
calculated from the simulation (square symbols; the solid line is
a guide to the eye), or
using the Gaussian theory (purple dashed-dotted line going to zero at long times). 
In dashed line (red) the Bogoliubov prediction.
In dashed-dotted-dotted line (blue)
the long time prediction (\ref{eq:pdc_ergo}) of the ergodic theory.
The parameters $T$, $N$ and $\rho g$ have the same values
as in Fig.\ref{fig:16a0}.}
\label{fig:corrthdot}
\end{figure}

\noindent{\it Gaussian theory}:
If we add by hand a decorrelation of the $b_{\bf k}$'s and assume 
Gaussian statistics, 
we get a diffusive spreading of the condensate phase change, 
with the variance of $\varphi_{\rm non\ osc}$ growing linearly at long times:
\begin{equation}
(\mbox{Var}\,\varphi)_{\rm Gauss} = 
\left(\frac{g}{\hbar V}\right)^2 \sum_{\mathbf{k}\neq\mathbf{0}}
(\tilde{U}_k+\tilde{V}_k)^4 \, \tilde{n}_k^2 \, \left[ \frac{e^{-2\gamma_k t}-1}{2 \gamma_k^2} + 
\frac{t}{\gamma_k}\right] \,,
\label{eq:DampedBogvarphase}
\end{equation}
in clear contradiction with the numerical simulations.
This prediction corresponds to a correlation function $\mathcal{C}$
vanishing at long times, whereas the correct correlation function 
has a finite limit, see Fig.\ref{fig:corrthdot}.

\noindent {\it Ergodic theory:} as in subsection \ref{subsec:corr_n0}
we calculate the long time value of the correlation function for
$\dot\varphi$ using the ergodic assumption. The various steps of the calculation
are rigorously the same as in Sec.~\ref{subsec:corr_n0} and lead to
\begin{equation}
\mathcal{C}(\tau\rightarrow +\infty)_{\rm ergo} =
\left(\frac{g}{\hbar V}\right)^2
\frac{1}{\mathcal{M}} \left(\sum_{\mathbf{k}\neq \mathbf{0}}
(\tilde{U}_k+\tilde{V}_k)^2 \tilde{n}_k\right)^2.
\label{eq:pdc_ergo}
\end{equation}
This prediction is in excellent agreement with the simulations:
it gives the correct asymptotic value of $\mathcal{C}$,
see Fig.\ref{fig:corrthdot}, and from the asymptotic
expression $\mbox{Var}\, \varphi \simeq \mathcal{C}(+\infty)
t^2$ it gives the correct values of the long time limit
of $(\mbox{Var}\,\varphi)^{1/2}/t$,
see Fig.\ref{fig:tphi}a, as a function of temperature.

\section{Quantum Treatment: Analytical Results} 
\label{sec:qtar}

So far the classical field model was very useful in revealing the physical
processes governing the long time behavior of the phase and atom
number fluctuations in the condensate. However it is not
a fully quantitative theory, as the long time limits of the
correlation functions considered here depend on the precise choice
of the energy cut-off, that is on the number of Bogoliubov modes $\mathcal{M}$
in the simulation, as is apparent on 
Eqs.(\ref{eq:dn0c_ergo},\ref{eq:pdc_ergo}). In this section, we therefore adapt
the previous physical reasonings to the quantum field case.

\subsection{The quantum model}

We use a straightforward generalization of the classical field lattice
model, taking here for simplicity a cubic lattice, as discussed in \cite{les_houches,Cartago,Mora}.
The bosonic field $\hat{\psi}$ evolves according to the Hamiltonian
\begin{equation}
H = \sum_{\mathbf k} \frac{\hbar^2 k^2}{2m} \, \hat{a}_\mathbf{k}^\dagger
\hat{a}_\mathbf{k} + \frac{g_0}{2} \sum_{\mathbf{r}} dV\,
\hat{\psi}^\dagger \hat{\psi}^\dagger \hat{\psi} \hat{\psi},
\end{equation}
where $\hat{a}_\mathbf{k}$ annihilates a particle of wavevector
$\mathbf{k}$ in the first Brillouin zone.
The dispersion relation of the wave is now the usual one.
The total number of atoms is fixed, equal to $N$.
The coupling constant $g_0$ depends on the lattice spacing $l$
in order to ensure a $l$-independent scattering length for the discrete delta interaction potential
among the particles
\cite{Mora,les_houches}: 
\begin{equation}
V(\mathbf{r}_1-\mathbf{r}_2) = \frac{g_0}{dV} \delta_{\mathbf{r}_1,\mathbf{r}_2}.
\end{equation}
Since we consider here
the weakly interacting regime, we can restrict
to a lattice spacing much larger than the scattering length
$a$ so that $g_0$ is actually very close to $g=4\pi \hbar^2 a/m$.

To be able to use Bogoliubov theory as we did
in the classical field reasoning, we restrict to the low
temperature regime $T\ll T_c$ with a macroscopic occupation of the condensate
mode. We thus neglect the possibility that the condensate is empty,
which allows us to use the modulus-phase representation of the condensate mode:
\begin{equation}
\hat{a}_0 \simeq e^{i\hat{\theta}} \sqrt{\hat{N}_0}
\end{equation}
where $\hat{N}_0 = \hat{a}_0^\dagger \hat{a}_0$
and where $\hat{\theta}$ is a Hermitian `phase' operator obeying
the commutation relation
\begin{equation}
[\hat{N}_0, \hat{\theta}] = i.
\label{eq:comm}
\end{equation}
This allows to consider the correlation of the condensate atom
number fluctuation $\delta\hat{N}_0\equiv \hat{N}_0-
\langle \hat{N}_0\rangle$ but also the variance of the condensate phase
change $\hat{\varphi}(t) \equiv \hat{\theta}(t)-\hat{\theta}(0)$,
as we did for the classical field.

\subsection{Correlation function of the condensate atom number}
\label{sub:quant_corr_n0}

To predict the correlation function
of $\delta\hat{N}_0$, we use Bogoliubov theory at short
times and the quantum analog of the ergodic theory at long times.

In the number conserving Bogoliubov theory \cite{Gardiner,CastinDum}, written
here for a spatially homogeneous system, one introduces the field conserving the
total number of particles
\begin{equation}
\hat{\Lambda}(\mathbf{r}) \equiv e^{-i\hat{\theta}} \hat{\psi}_\perp(\mathbf{r})
\end{equation}
where the non-condensed field $\hat{\psi}_\perp$ is obtained by projecting
out the component of the field $\hat{\psi}$ on the condensate mode.
The field $\hat{\Lambda}$ then admits the modal expansion on the Bogoliubov
modes
\begin{equation}
\hat{\Lambda}(\mathbf{r})  = \sum_{\mathbf{k}\neq\mathbf{0}}
\hat{b}_\mathbf{k} U_k \frac{e^{i\mathbf{k}\cdot\mathbf{r}}}{\sqrt{V}}+
\hat{b}^\dagger_\mathbf{k} V_k \frac{e^{-i\mathbf{k}\cdot\mathbf{r}}}{\sqrt{V}}
\label{eq:over_Bog}
\end{equation}
where the real amplitudes $U_k$, $V_k$, normalized as $U_k^2-V_k^2=1$,
are given by the usual Bogoliubov theory,
\begin{equation}
U_k+V_k = \left(\frac{\hbar^2 k^2/2m}{2\rho g_0+\hbar^2 k^2/2m}\right)^{1/4}.
\end{equation}

Since the total number of particles is fixed to $N$, it is equivalent to
consider the fluctuations of $\hat{N}_0$ or of the number of non-condensed
atoms
\begin{equation}
\hat{N}_\perp = \sum_\mathbf{r} dV \hat{\Lambda}^\dagger(\mathbf{r}) 
\hat{\Lambda}(\mathbf{r}).
\end{equation}
This, together with the expansion (\ref{eq:over_Bog}), expresses
$\hat{N}_\perp$ as a function of the $\hat{b}_\mathbf{k}$'s.

The equilibrium state of the system is approximated in the canonical
ensemble by the Bogoliubov thermal density operator
\begin{equation}
\hat{\rho}_{\rm Bog}(T) = \frac{1}{Z_{\mathrm{Bog}}} e^{-\sum_{\mathbf{k}\neq\mathbf{0}}
\epsilon_k \hat{b}^\dagger_\mathbf{k} \hat{b}_\mathbf{k}/k_B T}
\label{eq:rho}
\end{equation}
where the normalization factor $Z_{\mathrm{Bog}}$ 
is the Bogoliubov approximation
for the partition function, and where we have introduced the Bogoliubov spectrum
\begin{equation}
\epsilon_k =\left[\frac{\hbar^2 k^2}{2m}
\left(\frac{\hbar^2 k^2}{2m}+2\rho g_0\right)\right]^{1/2}.
\end{equation}

\noindent {\it Bogoliubov theory}: 
In the Bogoliubov approximation for the time evolution, the $\hat{b}_\mathbf{k}$
merely accumulate a phase, at the frequency $\omega_k=\epsilon_k/\hbar$, similarly
to the classical field case. From Wick's theorem one then obtains
\begin{eqnarray}
\frac{1}{2}\langle \{\delta\hat{N}_0(t),\delta\hat{N}_0(0)\}\rangle_{\rm Bog}  =
\sum_{\bf k\neq\bf 0} \bar{n}_k (\bar{n}_k+1) (U_k^2+V_k^2)^2 \nonumber \\
 + 2 U_k^2 V_k^2 \cos(2\omega_k t) \, [\bar{n}_k^2+(\bar{n}_k+1)^2]
\label{eq:corr_n0_bq}
\end{eqnarray}
where 
\begin{equation}
\bar{n}_k(T)=\frac{1}{\exp(\epsilon_k/k_B T)-1}
\label{eq:nkb}
\end{equation}
is the mean occupation
number of the Bogoliubov mode $\mathbf{k}$. Note that we have considered here
the so-called symmetric correlation function (as $\{X,Y\}$ stands for the anticommutator
$XY+YX$ of two operators) which is a real quantity, equal to the real part of the non-symmetrized
correlation function.
The time coarse grained version of the prediction (\ref{eq:corr_n0_bq}) is obtained
by averaging out the oscillating terms, which amounts to considering
the correlation function of the temporally smoothed
operator number of non-condensed particles 
\begin{equation}
\hat{N}_\perp^{\rm non\ osc} \equiv 
\sum_{\mathbf{k}\neq\mathbf{0}}
\left[\left(U_k^2+V_k^2\right) \hat{b}_\mathbf{k}^\dagger \hat{b}_\mathbf{k}+V_k^2\right].
\label{eq:nperp_no}
\end{equation}

\noindent {\it Quantum Ergodic theory:} 
Discarding from the start the oscillating terms in $\hat{N}_\perp$, 
as in (\ref{eq:nperp_no}), we face here the problem
of calculating the long time limit of $\langle A(t) A(0)\rangle$, where $A$ is a linear
function of the Bogoliubov mode occupation numbers,
\begin{equation}
A=\sum_{\mathbf{k}\neq\mathbf{0}} \gamma_k \hat{b}^\dagger_\mathbf{k} \hat{b}_\mathbf{k}\,.
\label{eq:defA}
\end{equation}
As the quantum state of the system is given by the Bogoliubov
approximation Eq.(\ref{eq:rho}), we may inject a closure relation in the Bogoliubov Fock eigenbasis:
\begin{eqnarray}
\langle A(t) A(0)\rangle &=& \frac{1}{Z_{\mathrm{Bog}}} \sum_{\{n_\mathbf{k}\}} 
e^{-\beta \sum_{\mathbf{k}\neq\mathbf{0}}\epsilon_k n_\mathbf{k}}
\left(\sum_{\mathbf{k}\neq\mathbf{0}} \gamma_k n_\mathbf{k}\right) \nonumber \\
&& \times \langle \{n_\mathbf{k}\} | A(t) | \{n_\mathbf{k}\} \rangle,
\label{eq:clos}
\end{eqnarray}
where the sum is taken over all possible integer values of the occupation numbers, not to be confused
with the mean occupation numbers (\ref{eq:nkb}).

The non-explicit piece of this expression is the matrix element of $A(t)$, which may be reinterpreted
as follows:
\begin{equation}
\langle \{n_\mathbf{k}\} | A(t) | \{n_\mathbf{k}\} \rangle = \mbox{Tr}\, \left[A \sigma(t)\right]
\label{eq:reint}
\end{equation}
where the density operator $\sigma$, initially a pure state in the 
Bogoliubov Fock basis,
\begin{equation}
\sigma(0) = |\{n_\mathbf{k}\} \rangle \langle \{n_\mathbf{k}\} |
\end{equation}
evolves during $t$ with the full Hamiltonian $H$. We know that this evolution involves Beliaev-Landau
processes that will spread $\sigma$ over the various Fock states $|\{n'_\mathbf{k}\} \rangle$.
This evolution is complex.
But we need here the long time limit only, in which we may assume
that an equilibrium statistical description is possible.
Since the system is isolated during its evolution, we take for 
$\sigma(t\to+\infty)$ the equilibrium
density operator in the {\sl microcanonical} ensemble
\cite{Deutsch}, 
and we calculate the expectation value of $A$ with $\sigma(t\to+\infty)$ 
as we did for the classical field model. The  
calculation can be done in the thermodynamic limit.  As shown in the Appendix \ref{appen:a},
one can calculate to leading order in this limit the difference between canonical and microcanonical averages.

Here the microcanonical ensemble has an energy
$E=E_0^{\rm Bog}+\sum_{\mathbf{k}\neq \mathbf{0}} \epsilon_k n_\mathbf{k}$,
where $E_0^{\rm Bog}$ is the ground state Bogoliubov energy.
We introduce the effective temperature $T_{\rm eff}$ such that the mean energy
in the canonical ensemble at temperature $T_{\rm eff}$ is equal to $E$,
\begin{equation}
0=\langle H_{\rm Bog}\rangle (T_{\rm eff}) - E =
\sum_{\mathbf{k}\neq \mathbf{0}} \epsilon_k [\bar{n}_k(T_{\rm eff})-
n_{\mathbf{k}}]
\label{eq:tbs}
\end{equation}
where $\langle \ldots \rangle$ stands for an average in the canonical ensemble
and $H_{\rm Bog}$ is the Bogoliubov Hamiltonian.
Using the results of Appendix \ref{appen:a} one gets
\begin{equation}
\bar{A}(E) - \langle A\rangle(T_{\rm eff}) =
-\frac{1}{2} k_B T_{\rm eff}^2 \left(\frac{\langle A\rangle'}{\langle H_{\rm Bog}\rangle'}\right)'(T_{\rm eff})
\label{eq:res2}
\end{equation}
where $\bar{A}(E)$ is the microcanonical average of $A$ at energy $E$ and
where the apex $'$ stands for derivation with respect to temperature.
We further use the fact that, in the thermodynamic limit, for typical values of the occupation
numbers $n_\mathbf{k}$, $T_{\rm eff}$ weakly deviates from the physical temperature $T$. 
We calculate $T_{\rm eff}$ by expanding (\ref{eq:tbs}) up to second order in 
$T_{\rm eff}-T$
\cite{more_details}. Evaluating (\ref{eq:res2}) with this value of $T_{\rm eff}$, keeping terms
up to the relevant order \cite{more_details}, gives the desired result
\begin{eqnarray}
&&\langle \{n_\mathbf{k}\} | A(t\rightarrow +\infty) | \{n_\mathbf{k}\} \rangle 
= \langle A\rangle  \nonumber \\
&+&\left[\sum_{\mathbf{k}\neq\mathbf{0}} \epsilon_k (n_\mathbf{k}-\bar{n}_k)\right]
\frac{\langle A\rangle'}{\langle H_{\rm Bog}\rangle'} +
\frac{1}{2} \left(\frac{\langle A\rangle'}{\langle H_{\rm Bog}\rangle'}\right)'
 \nonumber \\
&\times& \left\{
\frac{\left[\sum_{\mathbf{k}\neq\mathbf{0}} \epsilon_k (n_\mathbf{k}-\bar{n}_k)\right]^2}
{\langle H_{\rm Bog}\rangle'}-k_B T^2
\right\}
\label{eq:the_result}
\end{eqnarray}
where all the canonical averages are now evaluated at the physical temperature $T$ \cite{check}.

It remains to inject this expression into Eq.(\ref{eq:clos}). The resulting average over $n_{\bf k}$ leads to
the long time value of the correlation function:
\begin{eqnarray}
\langle A(+\infty) A(0)\rangle -\langle A\rangle ^2 &=&
\left( \frac{\langle A \rangle^\prime}{\langle  H_{\rm Bog} \rangle^\prime}\right)^2 
{\rm Var}\, H_{\rm Bog} \label{eq:fluctH} \\ &=&
\frac{\left[\sum_{\mathbf{k}\neq\mathbf{0}} \gamma_k \epsilon_k \bar{n}_k (\bar{n}_k+1)\right]^2}
{\sum_{\mathbf{k}\neq\mathbf{0}} \epsilon_k^2 \bar{n}_k (\bar{n}_k+1)},
\label{eq:caergo}
\end{eqnarray}
where we used Wick theorem and
the property $d\bar{n}_k/dT = \epsilon_k \bar{n}_k (\bar{n}_k +1)/ k_B T^2$ 
\cite{troisieme}.
Using Schwartz inequality, 
one can show that this long time value of the correlation function
is less than its zero time value $\sum \gamma_k^2 \bar{n}_k(\bar{n}_k+1)$.
To be complete,
we present an alternative derivation of our prediction (\ref{eq:caergo})
in the Appendix \ref{appen:alter}, based on results obtained in \cite{Deutsch}.
We also note that the quantum ergodic calculation directly leads to a prediction
of the long time limit for the correlation function
of the Bogoliubov mode occupation numbers, see Eq.(\ref{eq:qdndn}).

Replacing in Eq.(\ref{eq:caergo}) the coefficients $\gamma_k$ by their expression from Eq.(\ref{eq:nperp_no}),
$\gamma_k=U_k^2+V_k^2$, we obtain
the long time value of the condensate atom number correlation function 
in the quantum ergodic theory. Note that, in the thermodynamic limit, 
this long time value scales as the volume $V$, whereas the $t=0$ value scales as
$V^{4/3}$.

\subsection{Correlation function of the time derivative of the condensate
phase}

As in the classical field case, we first look for an expression of the first order time derivative
of the condensate phase operator $\hat{\theta}$ in terms of the amplitudes of the field $\hat{\Lambda}$ on
the Bogoliubov modes. Taking as a starting point in Heisenberg picture
\begin{equation}
i\hbar \frac{d}{dt} \hat{\theta} = [\hat{\theta},H],
\end{equation}
we split the quantum field in a condensate part and a non-condensed part,
\begin{equation}
\hat{\psi}(\mathbf{r}) = \frac{\hat{a}_0}{\sqrt{V}} + \hat{\psi}_\perp(\mathbf{r}),
\end{equation}
and we insert this splitting in the expression of $H$. 
Using the modulus-phase representation
of $\hat{a}_0$ and the commutation relation Eq.(\ref{eq:comm}), we obtain,
using $\hat{a}_0^\dagger \hat{a}_0 = \hat{N} - \hat{N}_\perp$,
\begin{eqnarray}
-\hbar\frac{d}{dt}\hat{\theta} &=& \frac{g_0}{V} \left[\hat{N}-\frac{1}{2} + 
\sum_{\mathbf{r}}  dV \hat{\Lambda}^{\dagger} \hat{\Lambda}\right] \nonumber \\
&+& \frac{g_0}{2V} \sum_\mathbf{r}  dV \left[ \hat{\Lambda} \frac{\hat{N}_0+1/2}{\sqrt{\hat{N}_0(\hat{N}_0+1)}}
\hat{\Lambda}+\mbox{h.c.} \right] \nonumber \\
&+& \frac{g_0}{2\sqrt{V}} \sum_\mathbf{r} dV\, \left[\frac{1}{\sqrt{\hat{N}_0}}\,\hat{\Lambda}^\dagger \hat{\Lambda}^2+\mbox{h.c.}\right].
\end{eqnarray}

The quantity $(\hat{N}_0+1/2)/\sqrt{\hat{N}_0(\hat{N}_0+1)}$ is actually $1+O(1/\hat{N}_0^2)$ so it can to a high accuracy 
be replaced by unity. 
Furthermore, as we did in the classical field model, we now 
keep the leading terms in $\hat{\Lambda}$, under the assumption
of a weak non-condensed fraction. We can also replace $\hat{\theta}$ by $\hat{\varphi}$ under the temporal derivative,
since $\hat{\theta}(0)$ is time independent. We obtain \cite{notN0}
\begin{equation}
-\hbar \frac{d}{dt}\hat{\varphi} \simeq \frac{g_0}{V} \left[\hat{N}-\frac{1}{2}+ \sum_{\mathbf{r}}  dV 
\left(\hat{\Lambda}^{\dagger} \hat{\Lambda} + \frac{1}{2} \hat{\Lambda}^2+
\frac{1}{2} \hat{\Lambda}^{\dagger 2}\right)\right].
\label{eq:approx_bog_phi_dot}
\end{equation}

Taking the expectation value of this expression over the thermal state in the Bogoliubov approximation
leads to an expression coinciding 
with the value of the chemical potential predicted
by Eq.(103) of \cite{Mora}, 
which includes in a systematic way the first correction
to the pure condensate prediction $\rho g_0$ \cite{more_precisely}:
\begin{equation}
\mu(T) = \frac{g_0}{V}\left[N-\frac{1}{2}+\sum_{\mathbf{k}\neq \mathbf{0}}
 (U_k+V_k)^2 \bar{n}_k  + V_k(U_k+V_k)\right].
\label{eq:muT}
\end{equation}
At this order of the expansion, this analytically shows
that $-\hbar\langle d\hat{\theta}/dt\rangle$ is the chemical potential of the system.
We now turn to various predictions for
the symmetrized correlation function of $d\hat{\varphi}/dt$,
\begin{equation}
\mathcal{C}_{\rm S}(\tau) = \frac{1}{2} \Big\langle\left\{\left(\frac{d}{dt}\hat{\varphi}\right)(\tau),
\left(\frac{d}{dt}\hat{\varphi}\right)(0)\right\}\Big\rangle - \langle\frac{d}{dt}\hat{\varphi}\rangle^2.
\end{equation}

\noindent {\it Bogoliubov theory:} At a time short enough for the interactions between
the Bogoliubov modes to remain negligible, one can apply Bogoliubov theory to get
\begin{eqnarray}
\mathcal{C}_{\rm S}^{\rm Bog}(\tau) & =&
\left(\frac{g_0}{\hbar V}\right)^2
\sum_{\mathbf{k}\neq\mathbf{0}} (U_k+V_k)^4 \left\{  \bar{n}_k (\bar{n}_k +1) 
\phantom{\frac{a^b}{c^d}}\right. \nonumber \\
&+& \left. \frac{1}{2}\cos(2\omega_k t) \left[\bar{n}_k^2 + (\bar{n}_k+1)^2\right]\right\}.
\end{eqnarray}
The temporal coarse grained version of this correlation function is obtained
by averaging out the cosine terms, which amounts to considering 
a temporal derivative of $\hat{\varphi}$ freed from the oscillating terms
$\hat{b}\hat{b}$ and $\hat{b}^\dagger \hat{b}^\dagger$:
\begin{eqnarray}
\left(\frac{d}{dt}\hat{\varphi}\right)_{\rm non osc} &=&
-\frac{g_0}{\hbar V} \left[\hat{N}-\frac{1}{2} + \sum_{\mathbf{k}\neq\mathbf{0}} V_k (U_k+V_k)\right] 
\nonumber 
 \\
&-& \frac{g_0}{\hbar V} \sum_{\mathbf{k}\neq\mathbf{0}} (U_k+V_k)^2 \hat{b}^\dagger_\mathbf{k} 
\hat{b}_\mathbf{k}.
\label{eq:phid_no}
\end{eqnarray}

\noindent {\it Quantum ergodic theory:} We directly apply to the smoothed
temporal derivative (\ref{eq:phid_no}) the reasoning performed
in the previous subsection. 
Up to an additive constant,
Eq.(\ref{eq:phid_no}) is indeed of the form (\ref{eq:defA}),
with $\gamma_k = -(g_0/\hbar V) (U_k+V_k)^2$.
From (\ref{eq:caergo}) we therefore obtain the long time behavior
of the phase derivative correlation function
\begin{equation}
\mathcal{C}_{\rm S}^{\rm ergo}(+\infty) =
\left(\frac{g_0}{\hbar V}\right)^2
\frac{\left[\sum_{\mathbf{k}\neq\mathbf{0}} (U_k+V_k)^2 \epsilon_k 
\bar{n}_k (\bar{n}_k+1)\right]^2}
{\sum_{\mathbf{k}\neq\mathbf{0}} \epsilon_k^2 \bar{n}_k (\bar{n}_k+1)}.
\label{eq:qep}
\end{equation}
The long time limit of the variance of the phase difference is then \cite{comp_clas}
\begin{equation}
{\rm Var}\,\hat{\varphi} \sim \mathcal{C}_{\rm S}^{\rm ergo}(+\infty) t^2 \,.
\end{equation}
Although our conclusion of a ballistic behavior for the phase agrees
qualitatively with \cite{Kuklov}, the explicit expression of
the coefficient of $t^2$ differs from the one of \cite{Kuklov}
due the fact that we account for interactions among Bogoliubov modes 
such as the Beliaev-Landau processes leading to ergodicity in the system,
while in \cite{Kuklov} the many-body Hamiltonian is replaced by the Bogoliubov
Hamiltonian in the last stage of the calculation.
As can be seen from (\ref{eq:qep}) using Schwartz inequality, 
ergodicity results in a reduction of
phase fluctuations with respect to the Bogoliubov prediction. 

\begin{figure}
\includegraphics[width=8cm,clip=]{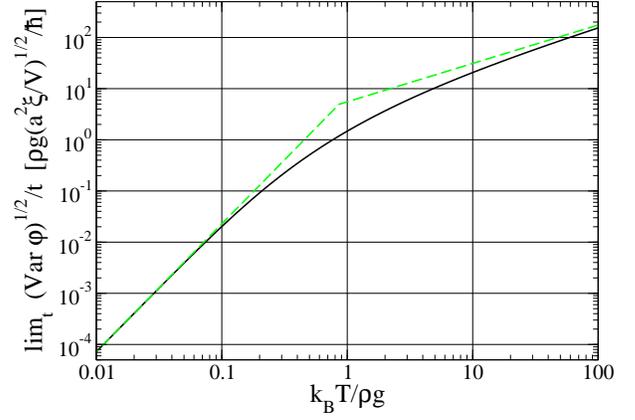}
\caption{(Color online) In the regime $T\ll T_c$ for a weakly interacting Bose gas,
quantum ergodic prediction (\ref{eq:qep}) for the long time limit
of $(\mbox{Var}\,\hat{\varphi})^{1/2}/t$, in the thermodynamic limit.
When expressed in units of $(a^2\xi/V)^{1/2} \rho g/\hbar$, 
$(\mbox{Var}\,\hat{\varphi})^{1/2}/t$ is a 
function of $k_B T/\rho g$ only, that is readily calculated numerically
(solid line) or that may be approximated by asymptotic
equivalents (\ref{eq:lowT},\ref{eq:highT}) in the low temperature or
high temperature limit (green dashed line).
Note that the dimensionless quantity $a^2\xi/V$ may also be written
as $\sqrt{\rho a^3}/(N\sqrt{4\pi})$.}
\label{fig:ergoq}
\end{figure}

In the thermodynamic limit, analytical expressions can be obtained
for this ergodic prediction. In the low temperature limit $k_B T \ll \rho g$,
\begin{equation}
\mathcal{C}_{\rm S}^{\rm ergo} (+\infty) \sim \frac{8\pi^4}{15} \frac{a^2 \xi}{V}
\left(\frac{k_B T}{\hbar}\right)^2 \left(\frac{k_B T}{\rho g}\right)^3,
\label{eq:lowT}
\end{equation}
where $\xi$ is the healing length such that $\hbar^2/m\xi^2 = \rho g$.
This tends to zero at zero temperature \cite{comp_to_Beliaev}.
In the high temperature limit $k_B T \gg \rho g$,
\begin{equation}
\mathcal{C}_{\rm S}^{\rm ergo} (+\infty) \sim \frac{12\zeta(3/2)^2}{5\zeta(5/2)}
\frac{a^2\lambda}{V} \left(\frac{k_B T}{\hbar}\right)^2  
\label{eq:highT}
\end{equation}
where the thermal de Broglie wavelength obeys $\lambda^2=2\pi\hbar^2/m k_B T$
and where $\zeta$ is the Riemann Zeta function.
Here we have identified $g_0$ to $g$ \cite{amusing}.
In Fig.\ref{fig:ergoq} we give the quantum ergodic prediction for
$\lim_{t\to\infty} (\mbox{Var}\,\hat{\varphi})^{1/2}/t$ calculated numerically,
which is
a universal function of $k_B T/\rho g$ when expressed in the right units. 

\section{Conclusion}
\label{sec:c}

We have investigated theoretically the phase spreading of a finite temperature
weakly interacting condensate. 
The gas is assumed to be prepared at thermal equilibrium in the
canonical ensemble, and then to freely evolve as an isolated system.
After average over many realizations of the system, we find in classical field
simulations that the variance $\mbox{Var}\,\varphi$ of
the condensate phase change grows quadratically in time. This 
non-diffusive behavior is quantitatively explained by an ergodic theory 
for the Bogoliubov modes,
the key point being that conservation of energy during the free
evolution prevents some correlation functions of the field from
vanishing at long times.
We have extended the analytical treatment to the quantum field case
and we have determined the coefficient of the $t^2$ term in the long
time behavior of $\mbox{Var}\,\hat{\varphi}$, see Eq.(\ref{eq:qep}).
This analytical result holds at low temperature $T\ll T_c$ and in the weakly interacting
regime $\rho a^3 \ll 1$, 
for a large number of thermally populated Bogoliubov modes,
and relies on the assumption that the (although weak) interaction among the Bogoliubov
modes efficiently mixes them (quantum ergodic regime). 

A physical insight in our result is obtained from the following rewriting 
\begin{equation} 
\mbox{Var}\,\hat{\varphi}(t) \sim \frac{t^2}{\hbar^2} 
\left(\frac{\partial_T\mu}{\partial_T \langle H\rangle}\right)^2 \mbox{Var}\,H
\label{eq:jolie}
\end{equation}
where $\mbox{Var}\,H$ is the variance of the energy of the gas,
here in the Bogoliubov approximation and in the canonical ensemble,
$\mu(T)$ is the chemical potential
of the system as given by Eq.(\ref{eq:muT})
and $\langle H\rangle(T)$ is its mean energy in the Bogoliubov approximation.

This formula may also be obtained from the following reasoning.
For a given realization of the system, of energy $E$,
the long time limit of the
condensate phase change $\hat{\varphi}(t)$ can be shown
to behave as
\begin{equation}
\hat{\varphi}(t) \sim -\mu_{\rm micro}(E) t/\hbar,
\label{eq:intuit}
\end{equation}
where $\mu_{\rm micro}$ 
is the chemical potential calculated in the microcanonical ensemble \cite{proof}.
For a large system, canonical energy fluctuations around the mean energy $\langle H\rangle(T)$ are weak
in relative value
so that one may expand $\mu_{\rm micro}(E)$ to first order in $E-\langle H\rangle$.
Taking the variance of $\hat{\varphi(t)}$ over the canonical fluctuations of
$E$ then leads to (\ref{eq:jolie}), since
$\partial_T\mu/\partial_T \langle H\rangle 
\simeq \partial_E \mu_{\rm micro} (\langle H\rangle)$
for a large system.

This reasoning shows that a necessary condition for the observation
of an {\it intrinsic} diffusive spreading of the condensate phase change 
is a strong suppression
of the energy fluctuations of the gas. To this end one may try to prepare the system in 
a clever way, starting with a pure condensate and giving to the system a well
defined amount of energy, e.g.\ by a reproducible change
of the trapping potential \cite{Thomas}. Alternatively one may try to follow 
a given experimental realization of the system, measuring the phase of the condensate 
in a non-destructive way and replacing ensemble average by time average.

We thank Fabrice Gerbier and Li Yun for useful comments on the
manuscript, and Mikhail Kolobov for interest in the problem. 
We thank Francis Hulin-Hubard for giving us access to a multiprocessor
machine.
We are grateful to Anatoly Kuklov for pointing to us his work \cite{Kuklov}
and for useful discussions.
One of us (E. W.) acknowledges financial support from QuFAR.
Laboratoire Kastler Brossel is a research unit of University Paris 6 and Ecole
normale sup\'erieure, associated to CNRS. Our group is a member of IFRAF.

\appendix

\section{Temporal correlation function of the Bogoliubov mode occupation numbers}
\label{appen:pilote}

Using the master equation approach developed in quantum
optics \cite{Louisell,CCT}, we calculate the temporal correlation
function of the operator $\hat{b}_\mathbf{q}^\dagger \hat{b}_\mathbf{q}$
giving the number of Bogoliubov excitations in the mode of wavevector
$\mathbf{q}$, in the thermodynamic limit and including the Beliaev-Landau
coupling among the Bogoliubov modes.
This is useful
to motivate the Gaussian model introduced
in section \ref{sec:cfar}, and to estimate the time
required for the correlation function 
$\langle A(t) A(0)\rangle -\langle A\rangle^2$, where $A$ is of the form
Eq.(\ref{eq:defA}), to depart from its value predicted by the Bogoliubov 
theory.

The idea of the master equation approach is to split the whole system
in a small system $S$ and
a large reservoir $R$ with a continuous energy spectrum.
Treating the coupling $W$ between $S$ and $R$ in the Born-Markov approximation
one obtains a master equation for the density operator $\sigma_S$
of the small system. Here the small system is the considered Bogoliubov
mode, with unperturbed Hamiltonian $H_S= \epsilon_q 
\hat{b}_\mathbf{q}^\dagger \hat{b}_\mathbf{q}$, and the reservoir is the set
of all other Bogoliubov modes, with unperturbed Hamiltonian 
$H_R= H_{\rm Bog} - H_S$. In the thermodynamic limit, the reservoir
indeed has a continuous spectrum, whereas the small system has a discrete
spectrum. The coupling $W$ between $S$ and $R$ is obtained from the
next order Bogoliubov expansion of the Hamiltonian, that is from the
part of the Hamiltonian cubic in the field $\hat{\Lambda}$,
\begin{equation}
H_{\rm cub} = g_0 \rho^{1/2} \sum_{\mathbf{r}} dV\, 
\hat{\Lambda}^\dagger (\hat{\Lambda}+\hat{\Lambda}^\dagger) \hat{\Lambda}.
\end{equation}
Inserting the modal decomposition Eq.(\ref{eq:over_Bog}) in $H_{\rm cub}$, 
we isolate the terms
that are linear in $b_\mathbf{q}$ \cite{why_linear}:
\begin{eqnarray}
W &=& g \rho^{1/2} \left[\hat{b}_\mathbf{q} R^\dagger + \hat{b}_{\mathbf{q}}^\dagger R\right]\\
R &=& V^{-1/2} \sum_{\mathbf{k},\mathbf{k}'\neq \mathbf{0},\mathbf{q}}
\left[\delta_{-\mathbf{q},\mathbf{k}+\mathbf{k}'} \mathcal{A}_{k,k'}
\hat{b}_{\mathbf{k}}^\dagger \hat{b}_{\mathbf{k}'}^\dagger\right.
\nonumber \\
&+&\left. \delta_{\mathbf{q},\mathbf{k}+\mathbf{k}'} \mathcal{B}_{k,k'}
\hat{b}_{\mathbf{k}} \hat{b}_{\mathbf{k}'} 
+ \delta_{\mathbf{q},\mathbf{k}'-\mathbf{k}} 2\mathcal{C}_{k,k'}
\hat{b}_{\mathbf{k}}^\dagger \hat{b}_{\mathbf{k}'} \right]
\end{eqnarray}
where the operator $R$ acts on the reservoir only and the coefficients
have the explicit expressions:
\begin{eqnarray}
\mathcal{A}_{k,k'} = U_q V_{k} V_{k'} + (U_q+V_q)(U_{k} V_{k'} +U_{k'} V_{k}) + V_q U_{k} U_{k'}
\nonumber\\
\mathcal{B}_{k,k'} = U_q U_{k} U_{k'} + (U_q+V_q)(V_{k} U_{k'} +U_{k} V_{k'}) + V_q V_{k} V_{k'}
\nonumber\\
\mathcal{C}_{k,k'} = U_q V_{k} U_{k'} + (U_q+V_q)(U_{k} U_{k'} +V_{k} V_{k'}) + V_q U_{k} V_{k'}.
\nonumber
\end{eqnarray}
As a consequence of momentum conservation for the whole system,
the action of $R$ (respectively $R^\dagger$)
changes the reservoir momentum by $-\hbar\mathbf{q}$ (respectively $\hbar\mathbf{q}$).

Let us denote with a tilde the operators in the interaction picture 
with respect to the Bogoliubov Hamiltonian $H_S+H_R$. 
In the Born-Markov approximation \cite{Louisell,CCT} 
the master equation, for the density operator of the small system in contact with the reservoir
in an equilibrium state, reads \cite{justif_bm}
\begin{equation}
\frac{d}{dt} \tilde{\sigma}_S(t) = 
-\int_0^{+\infty} \frac{d\tau}{\hbar^2}\,
\mbox{Tr}_R \left\{[\tilde{W}(t),[\tilde{W}(t-\tau),\tilde{\sigma}_S(t)
\sigma_R^{\rm eq} ]]\right\}
\end{equation}
where $\mbox{Tr}_R$ denotes the trace over the modes of the reservoir
and the equilibrium density operator of the reservoir
$\sigma_R^{\rm eq}$ is
supposed here to be the Bogoliubov thermal equilibrium
at temperature $T$.
We expand the double commutator; 
because of momentum conservation, the resulting terms that contain
two factors $\tilde{b}_\mathbf{q}$ or two factors 
$\tilde{b}_\mathbf{q}^\dagger$ exactly vanish when one performs the corresponding
traces over the reservoir. 
Coming back to Schr\"odinger's picture we finally obtain
\begin{eqnarray}
\frac{d}{dt}\sigma_S &=&
\frac{1}{i\hbar} [\check{\epsilon}_q \hat{b}_\mathbf{q}^\dagger
\hat{b}_\mathbf{q}, \sigma_S]
+ \Gamma_{q}^{-} \hat{b}_\mathbf{q} \sigma_S \hat{b}_\mathbf{q}^\dagger
+ \Gamma_{q}^{+} \hat{b}_\mathbf{q}^\dagger \sigma_S \hat{b}_\mathbf{q}
\nonumber \\
&-&\frac{1}{2}\left\{\Gamma_{q}^{-} \hat{b}_{\mathbf{q}}^\dagger \hat{b}_{\mathbf{q}}
+\Gamma_{q}^{+} \hat{b}_\mathbf{q} \hat{b}_{\mathbf{q}}^\dagger,\sigma_S\right\},
\label{eq:me}
\end{eqnarray}
where $\{,\}$ is the anticommutator and the new mode frequency
is $\check{\epsilon}_q=\epsilon_q+\hbar \Delta_q$.
The effect of the reservoir
on the small system 
is then characterized by a frequency shift $\Delta_q$ of the mode,
whose explicit expression
we shall not need here \cite{involved}, and by two transition rates
$\Gamma_{q}^{+}$ and $\Gamma_{q}^{-}$ given by the Fourier transform of reservoir
correlation functions at the mode frequency:
\begin{eqnarray}
\Gamma_{q}^{+} &=& \frac{g_0^2\rho}{\hbar^2} \int_{-\infty}^{+\infty} d\tau\,
e^{-i\epsilon_q\tau/\hbar} \mbox{Tr}_R[\tilde{R}^\dagger(\tau)R\sigma_R^{\rm eq}] \\
\Gamma_{q}^{-} &=& \frac{g_0^2\rho}{\hbar^2} \int_{-\infty}^{+\infty} d\tau\,
e^{i\epsilon_q\tau/\hbar} \mbox{Tr}_R[\tilde{R}(\tau)R^\dagger\sigma_R^{\rm eq}].
\end{eqnarray}
Since the reservoir is here at thermal equilibrium, the two rates are not
independent but $\Gamma_{q}^{-}=e^{\beta\epsilon_q} \Gamma_{q}^{+}$. This
results from the Bose law property $1+\bar{n}_k= e^{\beta\epsilon_k} \bar{n}_k$. The rates
are then conveniently characterized by their difference
$\Gamma_q \equiv \Gamma_{q}^{-}-\Gamma_{q}^{+}$. One finds
\begin{eqnarray}
\Gamma_q&=&\frac{g_0^2\rho}{(2\pi)^2\hbar} \int d^3 k
\left[
4\mathcal{C}_{k,k'}^2 (\bar{n}_k-\bar{n}_{k'})
\delta(\epsilon_q+\epsilon_k-\epsilon_{k'})
\right.  \nonumber \\
&&\left.+
2\mathcal{B}_{k,k'}^2 (1+\bar{n}_k+\bar{n}_{k'}) 
\delta(\epsilon_k+\epsilon_{k'}-\epsilon_q)
\right]
\end{eqnarray}
where $k'$ stands for $|\mathbf{k}-\mathbf{q}|$ in the integrand.
From \cite{Giorgini} one checks that $\Gamma_q$ is simply the standard
Beliaev-Landau damping rate for the Bogoliubov mode $\mathbf{q}$,
the contribution in $\mathcal{C}^2$ corresponding to the Landau mechanism
and the one in $\mathcal{B}^2$ to the Beliaev mechanism.

We now proceed with the calculation of the temporal correlation function 
of two operators $A_S, B_S$ of the small system, the whole system 
being at thermal
equilibrium. The quantum regression theorem \cite{Lax} states that
\begin{equation}
\langle A_S(t) B_S\rangle = \langle\langle A_S\rangle\rangle(t)
\equiv \mbox{Tr}_S\left[A_S \sigma_S^{\rm eff}(t)\right]
\end{equation}
for $t\geq 0$,
where the effective density operator $\sigma_S^{\rm eff}$ is in general
not hermitian nor of unit trace but evolves with the same master equation
as $\sigma_S$ with the initial condition
\begin{equation}
\sigma_S^{\rm eff}(0) = B_S \sigma_S^{\rm eq}
\end{equation}
where $\sigma_S^{\rm eq}=e^{-\beta H_S}/Z_S$ is the unit trace equilibrium 
solution of Eq.(\ref{eq:me}).
Using the invariance of the trace under a cyclic permutation we obtain
\begin{eqnarray}
\frac{d}{dt}\langle\langle A_S\rangle\rangle &=&
\frac{i\check{\epsilon}_q}{\hbar} \langle\langle [\hat{b}_\mathbf{q}^\dagger \hat{b}_\mathbf{q},
A_S]\rangle\rangle \nonumber \\
&+&\frac{\Gamma_{q}^{-}}{2} \langle\langle [\hat{b}_\mathbf{q}^\dagger,A_S]\hat{b}_\mathbf{q}
+\hat{b}_\mathbf{q}^\dagger[A_S,\hat{b}_\mathbf{q}] \rangle\rangle
\nonumber \\
&+&\frac{\Gamma_{q}^{+}}{2} 
\langle\langle [\hat{b}_\mathbf{q},A_S]\hat{b}_\mathbf{q}^\dagger
+\hat{b}_\mathbf{q}[A_S,\hat{b}_\mathbf{q}^\dagger] \rangle\rangle.
\end{eqnarray}
Specializing to $A_S=B_S^\dagger=\hat{b}_\mathbf{q}^\dagger$ or $\hat{b}_\mathbf{q}$
and $A_S=B_S=\hat{n}_\mathbf{q}\equiv \hat{b}_\mathbf{q}^\dagger
\hat{b}_\mathbf{q}$ leads to linear first order differential equations
for $\langle\langle A_S\rangle\rangle(t)$ that are readily solved:
\begin{eqnarray}
\langle b_\mathbf{q}^\dagger(t)b_\mathbf{q}\rangle &=& \bar{n}_q
e^{(i\check{\epsilon}_q-\Gamma_q/2)t}\\
\langle b_\mathbf{q}(t)b_\mathbf{q}^\dagger\rangle &=& (\bar{n}_q+1)
e^{(-i\check{\epsilon}_q-\Gamma_q/2)t}\\
\langle \hat{n}_\mathbf{q}(t) \hat{n}_\mathbf{q}\rangle
-\bar{n}_q^2 &=& \bar{n}_q (\bar{n}_q+1) e^{-\Gamma_q t}.
\label{eq:corrdiag}
\end{eqnarray}
In the classical field limit, where $\bar{n}_k+1$ is assimilated
to $\bar{n}_k$, this justifies the Gaussian theory of section \ref{sec:cfar}.
In both the classical and quantum cases, this shows that the occupation
numbers decorrelate with the rates $\Gamma_q$
corresponding to the Beliaev-Landau processes. These rates
have a non-zero value in the thermodynamic limit.

The present calculation is readily extended to the inclusion of two Bogoliubov
modes in the small system, of wavevectors $\mathbf{q}$
and $\mathbf{q}'\neq \mathbf{q}$. The coupling of the small system to the reservoir
now takes the form
\begin{equation}
W_2 = g \rho^{1/2} \left[\hat{b}_\mathbf{q} R_\mathbf{q}^\dagger + 
\hat{b}_{\mathbf{q}'}R_{\mathbf{q}'}^\dagger +\mbox{h.c.}\right],
\end{equation}
where the operators $R$ have the same structure as in the single mode
case, except that the double sum over $\mathbf{k},\mathbf{k}'$
is restricted to values different from $\mathbf{0},\mathbf{q},\mathbf{q}'$.
In the resulting master equation for the density operator of the two modes, 
the only issue is to see if there will be crossed terms between the two
modes, involving e.g.\ the product of $\hat{b}_\mathbf{q}^\dagger$ with 
$\hat{b}_{\mathbf{q}'}$. By calculating the trace over 
the reservoir of the corresponding product of operators $R$, e.g.\
$\mbox{Tr}_R[\tilde{R}_\mathbf{q}(\tau)\tilde{R}_{\mathbf{q}'}^\dagger\sigma_R^{\rm eq}]$,
we find in general that all crossed terms vanish, because of momentum conservation
\cite{exception}.
The master equation therefore does not couple the two modes,
and one obtains 
\begin{equation}
\langle \hat{n}_\mathbf{q}(t) \hat{n}_{\mathbf{q}'}\rangle
-\bar{n}_q \bar{n}_{q'} =0, \ \ \ \ \ \mbox{for}\ \mathbf{q}'\neq \mathbf{q},
\label{eq:corrcross}
\end{equation}
as is assumed in the Gaussian model for the classical field of section 
\ref{sec:cfar}.

It is instructive to compare the long time limit
of the predictions Eqs.(\ref{eq:corrdiag},
\ref{eq:corrcross}) to the quantum ergodic prediction.
Adapting the reasoning leading to Eq.(\ref{eq:caergo}), we obtain
the quantum ergodic result
\begin{equation}
\langle\hat{n}_\mathbf{q}(+\infty)\hat{n}_{\mathbf{q}'}\rangle
-\bar{n}_q \bar{n}_{q'} = \frac{\epsilon_q \epsilon_{q'} \bar{n}_q (\bar{n}_q+1)
\bar{n}_{q'}(\bar{n}_{q'}+1)}{\sum_{\mathbf{k}\neq\mathbf{0}} \epsilon_k^2 \bar{n}_k
(\bar{n}_k+1)}.
\label{eq:qdndn}
\end{equation}
In the thermodynamic limit this tends to zero, as in the master equation approach.

\section{Deviation of microcanonical and canonical averages}
\label{appen:a}

We wish to calculate the thermal expectation value of an observable $A$
in the microcanonical ensemble rather than in the canonical one.
For convenience, we shall parametrize the problem by the temperature $T$
of the canonical ensemble.
Restricting to the thermodynamic limit, where $k_B T$ is much larger than
the typical level spacing of the system, we calculate the first order
deviation of the two ensembles.

We start with the usual integral representation of the canonical ensemble in terms of the
microcanonical one:
\begin{equation}
\langle A\rangle (T) = \frac{\int dE\, \bar{A}(E) e^{S(E)/k_B} e^{-\beta E}}
{\int dE\, e^{S(E)/k_B} e^{-\beta E}}
\end{equation}
where the density of states is written in terms of the exponential of the microcanonical
entropy $S(E)$, $\bar{A}(E)$ and $\langle A\rangle(T)$
stand for the expectation value of $A$ 
in the microcanonical ensemble of energy $E$ and in the canonical ensemble 
of temperature $T$ respectively, and $\beta=1/k_B T$.

In the thermodynamic limit we expect the integrand to be strongly peaked around
the value $E_0(T)$ such that
\begin{equation}
\frac{d}{dE}\left[\frac{S(E)}{k_B}-\beta E\right]_{E=E_0(T)}=
S'[E_0(T)]/k_B - \beta=0,
\label{eq:implicit}
\end{equation}
where $f'(x)$ stands for the derivative of a function $f$ with respect to its argument $x$.
We then expand $u(E)\equiv S(E)/(k_B)-\beta E$ up to third order in $E-E_0$ and we
approximate the integrand as
\begin{eqnarray}
e^{u(E)} &=& e^{u(E_0)} e^{S''(E_0)(E-E_0)^2/2 k_B} \times \nonumber \\
&\times& \left(1 + \frac{1}{6} (E-E_0)^3 S^{(3)}(E_0)/k_B + \ldots \right).
\end{eqnarray}
We also expand $\bar{A}(E)$ up to second order in $E-E_0$.
Performing the resulting Gaussian integrals leads to
\begin{eqnarray}
&& \langle A\rangle (T) -  \bar{A}[E_0(T)] =
\frac{k_B}{2 |S''(E_0)|} \times \nonumber \\
&& \times  \left[ \bar{A}'(E_0) \frac{S^{(3)}(E_0)}{|S''(E_0)|}
+ \bar{A}''(E_0) \right]
+ \ldots
\label{eq:diff}
\end{eqnarray}
This relation can be inverted to first order, to give the microcanonical average
as a function of the canonical one; to this order,
we can assume that $\bar{A}[E_0(T)] = \langle A\rangle(T)$
in the right hand side of (\ref{eq:diff}).
Furthermore, using the implicit equation 
(\ref{eq:implicit}) one is able to express the derivatives with respect to $E_0$ 
in terms of derivatives with respect to $T$, e.g.\ $S''[E_0(T)]=-1/[T^2 E_0'(T)]$.
This leads to
\begin{equation}
\bar{A}[E_0(T)] - \langle A\rangle(T) =
-k_B T \left[\frac{\langle A\rangle'(T)}{E_0'(T)}+\frac{T \langle A\rangle''(T)}{2 E_0'(T)}\right]
+\ldots
\label{eq:res1}
\end{equation}

It is actually more convenient to parametrize the result in terms
of the mean canonical energy $\langle H\rangle(T)$ rather than
in terms of $E_0(T)$.
Applying (\ref{eq:res1}) to
$A=H$ allows to calculate
$E_0(T) - \langle H\rangle(T) $ to first order.
One then uses the first order expansion
\begin{eqnarray}
\bar{A}[\langle H\rangle(T)] &=&
\bar{A}[E_0(T)] + [\langle H\rangle(T)-E_0(T)]  \times \nonumber \\
&\times& \frac{1}{E_0'(T)} \frac{d}{dT}
\left\{\bar{A}[E_0(T)]\right\} + \ldots
\end{eqnarray}
In the first order term of this expression, 
we can replace $\bar{A}[E_0(T)]$ by the canonical 
average $\langle A\rangle(T)$, and we can identify
$E_0(T)$ with $\langle H\rangle(T)$; we can do the same identification
in the right hand side of (\ref{eq:res1}).
We obtain \cite{Weiss}
\begin{equation}
\bar{A}[\langle H\rangle(T)] - \langle A\rangle(T) =
-\frac{1}{2} k_B T^2 \frac{d}{dT}\left(\frac{d\langle A\rangle/dT}{d\langle H\rangle/dT}\right)
+\ldots
\end{equation}

\section{Alternative derivation of the long time limit of correlation functions}
\label{appen:alter}
We present in this section an alternative derivation of the ergodic result 
(\ref{eq:fluctH}) for the correlation function of an hermitian operator $A$,
here introduced in (\ref{eq:defA}). 
The long time limit of the correlation function is rigorously defined in terms
of the temporal average
\begin{equation}
C_{A}(+\infty) \equiv
\lim_{t\to +\infty} 
\frac{1}{t} \int_0^t \,d\tau \: \left[\langle A(\tau) A(0) \rangle
-\langle A\rangle^2\right].
\label{eq:timeave}
\end{equation}
We then insert in (\ref{eq:timeave}) a closure relation using the 
{\it exact} $N$-body eigenstates $|m\rangle$ of the interacting system with eigenenergies
$E_m$. In the absence of degeneracies we obtain a single sum over $m$,
\begin{equation}
C_A(+\infty) =
  \sum_m \, p_m \, \langle m|A|m \rangle^2  - 
  \left[\sum_m \, p_m \, \langle m|A|m \rangle \right]^2 \,.
\label{eq:sum_m}
\end{equation}
Here the $p_m=Z^{-1} \mbox{exp}(-\beta E_m)$ are the statistical 
weights defining the average in the canonical ensemble. 
Equation (\ref{eq:sum_m}), specialized for $\gamma_k=(g_0/V)(U_k+V_k)^2$,
is equivalent to Eq.(22) in \cite{Kuklov} 
for the dephasing time, provided one replaces there $H'$ by $A$. 
This makes the link between our approach and the one of \cite{Kuklov}.

The delicate point is now to relate the formal expression (\ref{eq:sum_m}) 
(involving the unknown exact eigenstates $|m\rangle$) to an explicit
expression treatable in the Bogoliubov approximation.
If one directly approximates the exact eigenstates by
eigenstates of the Bogoliubov Hamiltonian,
$|m\rangle \simeq |\{n_{\bf k}\}\rangle$, 
as done in \cite{Kuklov} (see Eq.(61) there),
one obtains the Bogoliubov result
\begin{equation}
C_A^{\rm Bog}(+\infty) = \sum \gamma_k^2 \bar{n}_k (\bar{n}_k +1),
\end{equation}
which is a good approximation for the $t=0$ value of the correlation function,
but not for its long time limit.
We argue that the exact eigenstates are in fact coherently spread over a large
number of Bogoliubov eigenstates of very close energies, because of the Beliaev-Landau
couplings among them.
Following \cite{Deutsch}, we thus assume that
\begin{equation} 
\langle m|A|m\rangle \simeq \bar{A}(E_m)
\end{equation} 
where $\bar{A}(E_m)$ is the microcanonical ensemble average at the energy $E_m$,
a thermodynamic quantity that is now treatable in the Bogoliubov approximation 
as we have already done in Eq.(\ref{eq:the_result}) \cite{Bog_non_Bog}.
After average over the canonical distribution for the energy $E_m=E$, 
we then obtain for the correlation function, 
\begin{equation}
C_A(+\infty) \simeq \langle \left[\bar{A}(E)-\langle A\rangle\right]^2\rangle
\simeq \left(\frac{\langle A\rangle'}{\langle H_{\rm Bog}\rangle'}\right)^2 
\mbox{Var}\, H_{\rm Bog},
\end{equation}
where $\langle \ldots\rangle$ stands for the canonical average at temperature $T$.
We recover Eq.(\ref{eq:fluctH}).


\begin{thebibliography}{99}
\bibitem{trick}
To get the equation defining $\tilde{E}_k$,
one would then replace, in Eq.(\ref{eq:Ektilde}), 
$\hbar^2 k^2/2m$ by $[(\hbar^2 k^2/2m) (2\rho g+\hbar^2 k^2/2m)]^{1/2}$ in the 
left hand side and 
$\tilde{E}_k$ by $[\tilde{E}_k (\tilde{E}_k+2\rho g)]^{1/2}$ in the right hand side.
\bibitem{Delta} $\Delta$ stands here for the operator acting on functions
on the lattice, such that its eigenvectors are the plane waves $e^{i\mathbf{k}\cdot
\mathbf{r}}$ with eigenvalues $-k^2$.
\bibitem{moment}
One may argue that the total momentum of the system provides three
additional constants of motion. Actually on the lattice model
considered here, the momentum along the direction $\alpha$
is conserved modulo $2\pi\hbar/l_\alpha$ only by the interaction term.
Since the energy cut-off in the simulations is of order of $k_B T$,
$2\pi\hbar/l_\alpha$ is also of the order of the typical momentum
of the populated modes of the field, so that we do not expect 
conservation of momentum. The simulations indeed confirm
this expectation.
\bibitem{notN0}
Note that in this equation the time derivative of the phase is not simply
proportional to $N_0$ or $\hat{N}_0$.
\bibitem{check}
As a consistency check one calculates the average of $\bar{A}$, that is of
Eq.(\ref{eq:the_result}), over the canonical
distribution for the occupation numbers $n_\mathbf{k}$: using
$d\bar{n}_k/dT = \epsilon_k \bar{n}_k (\bar{n}_k +1)/ k_B T^2$,
one indeed finds $\langle \bar{A} \rangle(T) = \langle A\rangle(T)$.
\bibitem{troisieme}
Due to the fact that the third term of Eq.(\ref{eq:the_result})
has a vanishing mean, its contribution
to $\langle A(+\infty)A(0) \rangle - \langle A \rangle^2$ 
is much smaller than the one of the second term
in the thermodynamic limit and can be neglected.
We then obtain Eq.(\ref{eq:caergo}).
\bibitem{more_precisely}
More precisely it exactly coincides if one replaces
the value of the chemical potential $\mu$ by its lowest order value $\rho g_0$
in the Bogoliubov expectation values of products of $\hat{\Lambda}$
 in the right hand side of Eq.(103) of \cite{Mora}.
This is allowed at the order of accuracy of Eq.(\ref{eq:approx_bog_phi_dot}).
\bibitem{comp_clas}
Taking the classical limit $\bar{n}_k+1\simeq \bar{n}_k \to \tilde{n}_k=k_BT/\tilde{\epsilon}_k$,
$U_k\to \tilde{U}_k$, $V_k\to \tilde{V}_k$, $\epsilon_k\to \tilde{\epsilon}_k$,
$g_0\to g$, we recover the classical prediction Eq.(\ref{eq:pdc_ergo}).
\bibitem{proof}
On one hand,
the microcanonical chemical potential is the derivative of the
energy with respect to $N$ for a fixed entropy, that is for a fixed number of
quantum states within the energy shell defining the microcanical ensemble.
In the Bogoliubov frame, we obtain, from the Hellmann-Feynman theorem,
$\mu_{\rm micro}= \partial_N E_0 + \sum_{\mathbf{k}\neq\mathbf{0}} 
\langle \hat{b}^\dagger_\mathbf{k} \hat{b}_\mathbf{k}\rangle_{\rm micro}
\partial_N \epsilon_k$, where $E_0=g_0 N(N-1)/2V 
-\sum_{\mathbf{k}\neq\mathbf{0}}
\epsilon_k V_k^2$ is the Bogoliubov ground state energy. We then calculate explicitly all the
derivatives with respect to $N$.
E.g., $\partial_N \epsilon_k = (g_0/V) (U_k + V_k)^2$.
On the other hand, in the spirit of the ergodic method
we calculate the expectation value
of $(d\hat{\varphi}/dt)_{\rm non\ osc}$ in the microcanonical ensemble of energy $E$.
The expressions of the chemical potential obtained in these two ways coincide.
Finally we split $(d\hat{\varphi}/dt)_{\rm non\ osc}$ as the sum of its mean 
microcanonical value $\bar{\dot{\varphi}}$ and of fluctuations.
Assuming that the fluctuations have vanishing correlations over long times for the considered
given realisation, we obtain Eq.(\ref{eq:intuit}) after temporal integration.
\bibitem{why_linear} Because of momentum conservation, $H_{\rm cub}$
has no cubic term in $b_\mathbf{q}$, and has a finite number of quadratic
terms in $b_\mathbf{q}$ which are thus negligible in the thermodynamic limit.
\bibitem{justif_bm} The Born-Markov approximation 
implies that the temporal correlation functions
of $\tilde{R}$ and $\tilde{R}^\dagger$ with $\tilde{R}$ and
$\tilde{R}^\dagger$ have a temporal width much smaller than
the damping time of $\tilde{\sigma}_R$. This is the case
in the Bogoliubov limit, as defined in \cite{CastinDum}, 
where $g_0 \to 0$ for fixed values of
 $\rho g_0$, $T$, $q$, $dV$, since the correlation functions
are then fixed and the damping rates $\Gamma_{\pm}$ 
tend to zero linearly in $g_0$.
\bibitem{involved} The correct calculation of $\Delta_q$
may involve the first order
effect of the quartic perturbation to $H_{\rm Bog}$, neglected here.
See \cite{Shlyapnikov} for a complete calculation.
\bibitem{exception}
The only exception is 
when $\mathbf{q}+\mathbf{q}'=\mathbf{0}$; the crossed terms
in this case, however, involve in the interaction picture
a product of $\tilde{b}_\mathbf{q}(t)$ and $\tilde{b}_{-\mathbf{q}}(t)$,
or a product of their hermitian conjugates; the corresponding functions of $t$
oscillate with the frequency $\pm 2\epsilon_q/\hbar$; in the Bogoliubov limit,
as in \cite{justif_bm},
this frequency is 
much larger than the relaxation rates of $\tilde{\sigma}_S(t)$, so that the
crossed terms can be neglected in the so-called secular approximation.
\bibitem{Weiss} If we apply this formula to the number of 
non-condensed particles in the 1D harmonically trapped ideal Bose gas,
we recover to first order equation (64) of \cite{Weiss2}.
\bibitem{Bog_non_Bog} In the weighted average of 
the matrix element $\langle m |A| m\rangle$, that is
$S_1=\sum_m p_m \langle m|A|m\rangle$,
the mixing of Bogoliubov states caused by Landau-Beliaev
processes does not show up: 
If one assumes that $\langle m|A|m\rangle$ is a microcanonical
average of Bogoliubov matrix elements of $A$,
one may reorder the terms in $S_1$, 
replacing the sum over the exact states by the sum over
the Bogoliubov states.
The same trick does not apply for the weighted average of the squared matrix
element $\langle m|A|m\rangle^2$.
The fluctuation properties of $\langle m|A|m\rangle$ 
are thus wrongly described if one replaces the exact eigenstates by the Bogoliubov ones,
unphysically large fluctuations being then introduced.
\end{thebibliography}
\end{document}